\pdfoutput=1
\documentclass[12pt]{iopart}

\usepackage{iopams}  
\expandafter\let\csname equation*\endcsname=\relax
\expandafter\let\csname endequation*\endcsname=\relax

\usepackage{physics}                            
\usepackage{bm}									
\usepackage{enumerate}							
\usepackage{amssymb}							
\usepackage[dvipsnames]{xcolor}								
\usepackage[colorlinks=true,allcolors=RoyalBlue]{hyperref}
\hypersetup{colorlinks=true}
\usepackage[capitalize]{cleveref}               
\usepackage{graphicx,color}
\usepackage{bbold}                              
\usepackage{cite}
\usepackage{tikz}
\usepackage[utf8]{inputenc}

\newcommand{\cor}[1]{\mathcal{#1}}									
\newcommand{\T}[1]{\text{#1}}										
\newcommand{\dslash}[1]{\frac{\dd[d]{#1}}{(2\pi)^d}}                
\newcommand{\dperp}[1]{\frac{\dd[d-1]{{#1}_\perp}}{(2\pi)^{d-1}}}
\def \z {^{(0)}}
\def \o {^{(1)}}
\def \t {^{(2)}}
\def \tr {^{(3)}}
\def \f {^{(4)}}
\def \i {^{(i)}}
\def \q {_{\vb q}}
\def \p {_{\vb p}}
\def \k {_{\vb k}}
\def \qp {_{\vb q,\vb p}}
\def \pq {_{\vb p,\vb q}}

\def \sc {^\T{(sc)}}

\newcommand{\n}{\nonumber}

\newcommand{\rev}[1]{#1}

\usepackage{etoolbox}
\makeatletter
\newrobustcmd{\fixappendix}{%
  \patchcmd{\l@section}{1.5em}{7em}{}{}%
  \patchcmd{\l@subsection}{2.3em}{7em}{}{}%
}
\makeatother

\newcommand{\thetitle}{Dynamics of soft interacting particles on a comb}

\begin{document}

\title[\thetitle]{\thetitle}

\author{Davide Venturelli$^{1,2}$, Pierre Illien$^{3}$, Aur\'elien Grabsch$^{1}$, and Olivier B\'enichou$^{1}$}

\address{$^1$ Laboratoire de Physique Th\'eorique de la Mati\`ere Condens\'ee, CNRS/Sorbonne Universit\'e, 4 Place Jussieu, 75005 Paris, France}
\address{$^2$ Laboratoire Jean Perrin, CNRS/Sorbonne Universit\'e, 4 Place Jussieu, 75005 Paris, France}
\address{$^3$ Laboratoire PHENIX (Physico-Chimie des Electrolytes et Nanosyst\`emes Interfaciaux), CNRS/Sorbonne Universit\'e, 4 Place Jussieu, 75005 Paris, France}
\ead{davide.venturelli@cnrs.fr}
\vspace{10pt}
\begin{indented}
\item[]\today
\end{indented}

\begin{abstract}
We study the dynamics of overdamped Brownian particles interacting through soft pairwise potentials on a comb-like structure. 
Within the linearized Dean--Kawasaki framework, we characterize the 
particle density fluctuations by computing their one- and two-point correlation functions. 
For a tracer particle constrained to move along the comb backbone, we determine the spatial correlation profile between its position and the density of surrounding bath particles.
Furthermore, we derive the correction to the diffusion coefficient of the tracer due to interactions with other particles, validating our results through numerical simulations.
\end{abstract}

%
%
%
%
%

\tableofcontents
\markboth{\thetitle}{\thetitle}

\section{Introduction}

Particles diffusing in complex environments can undergo anomalous diffusion, characterized by a nonlinear growth with time of their mean squared displacement $\expval{x^2(t)}$. 
The ubiquity of anomalous diffusion in soft and biological systems~\cite{Hofling:2013} has, over the past decades, stimulated significant interest in developing minimal theoretical models to capture this behavior~\cite{Metzler2000,Condamin2008}.
In this context, diffusion of particles in systems with geometrical constraints, such as fractal or disordered lattices, has
been the subject of an intense theoretical scrutiny~\cite{2005Havlin-book,Bouchaud1990}, also due to its relevance to the description of transport in porous media, polymer mixtures, and living cells.

Among the simplest inhomogeneous structures is the \textit{comb}: this can be visualized as a line (called the \textit{backbone}) spanning the system from one end to the other, and connected to 
infinite structures (the teeth), as depicted in \cref{fig:comb}.
Comb structures have been originally introduced to represent diffusion in critical percolation
clusters~\cite{White1984,Weiss1986,Havlin_1987}, with the backbone and teeth of the comb mimicking the quasi-linear structure and
dead ends of the clusters, respectively. 
In general, one expects that a particle will
spend a long time exploring a tooth, resulting in a sub-diffusive motion along the backbone --- e.g., $\expval{x^2(t)}\sim t^{1/2}$ in two spatial dimensions~\cite{Baskin_1991}. 
Since then, comb-like models have been used to describe real systems such as cancer proliferation~\cite{Iomin_2006} and dendronized polymers~\cite{Frauenrath2005}, transport in spiny dendrites~\cite{Mendez:2013}, and the diffusion of cold atoms~\cite{Sagi:2012} or in crowded media~\cite{Hofling:2013}.

From a theoretical perspective,
the single-particle comb model is sufficiently simple to allow for the derivation of several exact analytical results, both in the continuum~\cite{Arkhincheev2002,Baskin_2004,Lenzi_2013,Ribeiro2014,Iomin_2023,Trajanovski_2023},
and on the lattice~\cite{Burioni2005,Vulpiani_2011,Barkai_2013,Agliari_2014,Agliari_2015,Illien2016} --- see also Ref.~\cite{Iomin2018} for an overview. By contrast, the many-body problem, consisting of interacting particles evolving on a comb, has received much less attention, and so far restricted to the case of hard-core lattice gases~\cite{Benichou_2015,Poncet_2022,Grabsch:2023a,Berlioz2024}. The natural step forward, which we aim to address here, is to explore the dynamics of interacting particles in continuum space, subject to the comb constraint.

To this end, in this work we consider a system of overdamped Brownian particles interacting via soft pairwise potentials, and constrained to move on a comb, as described in \cref{sec:model}. 
Using the Dean--Kawasaki formalism~\cite{Dean1996,Kawasaki1998}, we first derive in \cref{sec:dean} the exact 
equations that describe the dynamics of the particle density field $\rho(\vb x,t)$.
Expanding the latter around a constant background density, in \cref{sec:density_dynamics} we then derive the one- and two-point functions that completely characterize the density fluctuations within the Gaussian approximation. 
To the best of our knowledge, this represents the first application of the Dean--Kawasaki theory to non-homogeneous media~\cite{teVrugt2020,illien2024deankawasaki}.
Next, in \cref{sec:tracer_dynamics}, we single out a tagged tracer from the bath of interacting particles, assuming that it is constrained to move only along the comb backbone.
In this setting, we first derive the spatial correlation profiles that describe the cross-correlations between the tracer position and the surrounding bath density, and then use them to estimate the effective diffusion coefficient of the tracer, which we finally test using Brownian dynamics simulations.

\begin{figure}
\centering
\begin{tikzpicture}[scale=0.9]
    \def\LineLength{3} 
    \def\Spacing{2}    
    \def\ParticleRadius{0.17} 
    \def\DashLength{0.2} 
    \def\Gap{0.1} 

    \draw[thick] (-0.5*\Spacing, 0) -- (5.5*\Spacing, 0);
    \draw[thick] (-0.5*\Spacing-\Gap, 0) -- (-0.5*\Spacing-\Gap-\DashLength, 0);
    \draw[thick] (5.5*\Spacing+\Gap, 0) -- (5.5*\Spacing+\Gap+\DashLength, 0);

    \foreach \x in {0, 1, 2, 3, 4, 5} {
        \draw[thick] (\x*\Spacing, -0.5*\LineLength) -- (\x*\Spacing, 0.5*\LineLength);
        \draw[thick] (\x*\Spacing, 0.5*\LineLength+\Gap) -- (\x*\Spacing, 0.5*\LineLength+\Gap+\DashLength);
        \draw[thick] (\x*\Spacing, -0.5*\LineLength-\Gap) -- (\x*\Spacing, -0.5*\LineLength-\Gap-\DashLength);
    }

    \fill[RoyalBlue] (0*\Spacing, 0.3*\LineLength) circle (\ParticleRadius);
    \fill[RoyalBlue] (0*\Spacing, -0.4*\LineLength) circle (\ParticleRadius);
    \fill[BrickRed] (1.2*\Spacing, 0) circle (\ParticleRadius);
    \fill[RoyalBlue] (2*\Spacing, 0.2*\LineLength) circle (\ParticleRadius);
    \fill[RoyalBlue] (3*\Spacing, -0.2*\LineLength) circle (\ParticleRadius);
    \fill[RoyalBlue] (3.75*\Spacing, 0*\LineLength) circle (\ParticleRadius);
    \fill[RoyalBlue] (4*\Spacing, -0.4*\LineLength) circle (\ParticleRadius);
    \fill[RoyalBlue] (5*\Spacing, 0.4*\LineLength) circle (\ParticleRadius);
    \fill[RoyalBlue] (5*\Spacing, 0.2*\LineLength) circle (\ParticleRadius);
\end{tikzpicture}
\caption{Schematic representation of interacting Brownian particles evolving on a comb structure in $d=2$. Bath particles (blue) can move horizontally along the \textit{backbone}, or vertically along the \textit{teeth}. Note that, in the continuum model considered here, the teeth are continuously distributed along the horizontal direction (in the sketch they are instead represented as spatially separated, only for graphical purposes). 
The tracer particle (red) is constrained to diffuse only along the backbone.}
\label{fig:comb}
\end{figure}

\section{The model}
\label{sec:model}

We consider a system of $(N+1)$ Brownian particles at positions $\vb r_i(t)$, which evolve according to the overdamped Langevin equations
\begin{equation}
    \dot{\vb r}_i(t)=
    -\hat \mu\i\left(\vb r_i(t)\right) \sum_{j\neq i} \nabla_{\vb r_i} U_{ij}\left(\vb r_i(t)-\vb r_j(t)\right)+\bm\eta_i\left(\vb r_i(t),t\right).
    \label{eq:langevin}
\end{equation}
\rev{Here, the Gaussian noises $\bm \eta_i$ satisfy $\expval{\bm \eta_i(\vb x,t)=\bm 0}$ and have variance}
\begin{equation}
    \left\langle \bm \eta_i(\vb x,t) \bm\eta^{\mathrm{T}}_j\left(\vb x',t^{\prime}\right)\right\rangle=2k_B T\hat \mu\i(\vb x) \delta_{i j} \delta(\vb x-\vb x')\delta\left(t-t^{\prime}\right) ,
    \label{eq:noise_particles}
\end{equation}
\rev{so that the fluctuation-dissipation theorem is satisfied (henceforth we will set the Boltzmann constant $k_B$ to unity, for simplicity).}
For future convenience, we single out the particle with $i=0$, which we denote here and henceforth as the \textit{tracer}. 
\rev{We allow for the particles' mobility to generically depend on their position in space, and we assume that the tracer can have a different mobility with respect to that of the other bath particles; hence, the mobility matrices  
in \cref{eq:langevin} read
\begin{equation}
    \hat \mu\i(\vb x) = 
    \begin{cases}
            \hat \mu\z (\vb x), & i=0,\\
        \hat \mu (\vb x), & i\geq 1.
    \end{cases}
\end{equation}
Finally, the inter-particle interaction potentials in \cref{eq:langevin} read
\begin{equation}
    U_{ij}(\vb x) = 
    \begin{cases}
    U_0(\vb x), & i=0 \T{ or } j=0 ,\\
        U(\vb x), & i,j\geq 1,        
    \end{cases}
    \label{eq:interaction_potential}
\end{equation}
between tracer-bath or bath-bath particles, respectively.
}

\rev{So far, the features of the underlying space in which the particles are evolving have not yet been specified. To enforce the \textit{comb} geometry, which is the subject of this work (see Fig.~\ref{fig:comb}), one can introduce in \cref{eq:langevin} the anisotropic mobility matrices~\footnote{We stress that, by dimensional consistency, the delta functions that appear in the expressions above should be replaced by $\delta(x_i)\mapsto a\, \delta(x_i)$, where $a$ is a length scale. In the following, we will set this length scale to unity for simplicity; indeed, note that $a$ can in any case be reabsorbed via a suitable rescaling of the corresponding Fourier variable $q_1$ (see c.f.~\cref{sec:density_dynamics}),~e.g.~$q_1^2\mapsto q_1^2 a^{d-1} $.\label{footnote-dimension}}}
\begin{equation}
    \hat \mu(\vb x) = \mu \mqty(\delta ( x_2)\dots \delta ( x_d) &0\\0&\mathbb{1}_{d-1})  = \mu \mqty(\delta ( \vb x_\perp) &0\\0&\mathbb{1}_{d-1}), 
    \quad \T{and} \quad \hat \mu\z(\vb x) = \mu_0 \mqty(1 &0\\0&\mathbb{0}_{d-1}).
    \label{eq:mobilities}
\end{equation}
This constrains the tracer particle to move only on the \textit{backbone} (which we chose without loss of generality to be oriented along the Cartesian direction $\vu e_1$), while all other bath particles can move along the backbone only if $\vb x_\perp=\bm 0$.
In $d=2$, \cref{eq:mobilities} reduces to
\begin{equation}
    \hat \mu(\vb x) = \mu \mqty(\delta ( y) &0\\0&1), \qquad \T{and} \qquad \hat \mu\z(\vb x) = \mu_0 \mqty(1 &0\\0&0),
    \label{eq:mobilities2d}
\end{equation}
corresponding to the comb schematically represented in \cref{fig:comb}. For $d\geq 3$, the resulting structure may be rather visualized as a backbone sliced by orthogonal hyperplanes.

Note that, for nontrivial choices of $\hat \mu\i(\vb x)$ in \cref{eq:noise_particles}, the Gaussian noises $\bm \eta_i$ are white and \textit{multiplicative} --- to fix their physical meaning, we adopt here the It\^o prescription.
Choosing $\hat \mu(\vb x)$ as in \cref{eq:mobilities2d}, it is then simple to check that the Langevin equation~\eqref{eq:langevin} reduces, in the single-particle case, to
\begin{equation}
    \partial_t P(\vb r,t) = \mu T \left[ 
    a 
    \delta(y) \partial_x^2 +\partial_y^2 \right] P(\vb r,t),
    \label{eq:single_comb}
\end{equation}
which is the Fokker-Planck equation for a Brownian particle at position $\vb r=(x,y)$ on a $2d$ comb, as commonly adopted in the literature~\cite{Baskin_1991,Iomin2018} (and where we have replaced $\delta (y)\mapsto a \delta(y)$, where $a$ is a length scale --- see the footnote on page~\pageref{footnote-dimension}).
From the well-known solution~\cite{Baskin_1991} of \cref{eq:single_comb} (recalled in \ref{app:density}, see c.f.~\cref{eq:MGF-free-part}), one finds at long times $t$ that $\expval{y^2(t)}\simeq 2\mu T t$, while $\expval{x^2(t)} \simeq 2a \sqrt{\mu T t/\pi }  $. Thus, a free Brownian particle on a comb diffuses along the teeth, but subdiffuses along the backbone.

Crucially, in the presence of interactions or external potentials~\cite{Trajanovski_2023},
inserting $\delta(y)$ only in the diffusion coefficient (as done in \cref{eq:single_comb} for the free-particle case) would not prevent the particles from leaving the comb. This is why the term $\delta(y)$ has been included also in the particle's mobility~\eqref{eq:mobilities}.

\rev{In the following section, we set out to compute, starting from the set of Langevin equations~\eqref{eq:langevin}, the evolution equation satisfied by the fluctuating particle density.}

\section{Joint dynamics of the tracer and bath particles density}
\label{sec:dean}
The dynamics of the
\rev{microscopic}
particle density can be derived by standard methods within the Dean--Kawasaki formalism, also known as stochastic density functional theory (SDFT)~\cite{Dean1996,Kawasaki1998,illien2024deankawasaki}. To this end, here we generalize the derivation of Ref.~\cite{Demery2014} to the case in which 
the particles' mobility encodes geometrical constraints, as in \cref{eq:mobilities}.
Following Refs.~\cite{Dean1996,Demery2014}, we first 
introduce the fluctuating density of bath particles
\begin{equation}
    \rho(\vb x,t) = \sum_{i=1}^N \delta \left( \vb x-\vb r_i(t) \right),
    \label{eq:density}
\end{equation}
in terms of which the evolution equation of the tracer \rev{(i.e.~$i=0$)} becomes
\begin{equation}
    \dot{\vb r}_0(t)= 
    -\hat\mu\z\left( \vb r_0(t)\right) \nabla_{\vb r_0} \int \dd{\vb y} U_0\left(\vb r_0(t)-\vb{y}\right)\rho(\vb y,t)+\bm\eta_0(\vb r_0(t),t),
\end{equation}
while the bath density can be shown to follow the Dean equation~\cite{Dean1996,Kawasaki1998}
\begin{equation}
    \partial_t \rho(\vb x,t) = \div \left[ \hat \mu(\vb x) \rho(\vb x,t) \grad \fdv{\mathcal F}{\rho(\vb x,t)} \right]+\div \left[ \rho^{\frac12}(\vb x,t) \bm \xi (\vb x,t) \right],
    \label{eq:dean}
\end{equation}
\rev{where $\bm \xi$ is a Gaussian noise field with zero mean and correlations}
\begin{equation}
    \left\langle \bm \xi(\vb x,t) \bm\xi^{\mathrm{T}}\left(\vb x',t^{\prime}\right)\right\rangle=2T\hat \mu(\vb x)\delta (\vb x-\vb x') \delta\left(t-t^{\prime}\right) ,
    \label{eq:xi_corr}
\end{equation}
\rev{and where $\mathcal F$ takes the form of a pseudo free energy,}
\begin{equation}
    \mathcal F[\rho,\vb r_0] = T \int \dd{\vb x}\rho(\vb x) \log( \frac{\rho(\vb x)}{\rho_0}) +  \frac 12 \int \dd{\vb x} \dd{\vb y} \rho(\vb x) U(\vb x-\vb y)\rho(\vb y) + \int \dd{\vb y}\rho(\vb y) U_0(\vb y-\vb r_0),
\end{equation}
\rev{with $\rho_0$ being a uniform background density.}

By choosing $U_0(\vb x-\vb y)=U_0(\vb y-\vb x)$ --- i.e.~by assuming that the 
\rev{interaction between the tracer and the other bath particles}
is \textit{reciprocal} --- then we may also rewrite 
\begin{equation}
    \dot{\vb r}_0(t)= 
    -\hat\mu\z\left( \vb r_0(t)\right) \nabla_{\vb r_0} \mathcal F[\rho,\vb r_0]+\bm\eta_0(\vb r_0(t),t),
    \label{eq:reciprocal}
\end{equation}
so that
the system 
formally admits the joint stationary distribution
\begin{equation}
    \cor P[\rho, \vb r_0] \propto \mathrm e^{-\frac{1}{T}F[\rho,\vb r_0]}.
    \label{eq:stationary_dist}
\end{equation}
Crucially, however, reaching this solution dynamically requires ergodicity, and this is not generically granted in the presence of geometrical constraints --- such as those encoded in $\hat \mu(\vb x)$ for the comb (see \cref{sec:comb}). 
\rev{Indeed, a}
simple counter-example can be produced by choosing $\hat \mu (\vb x) = \T{diag}(1,0)$ in $d=2$: clearly, the system cannot relax along the $\vu y$ direction, because there are no forces or noise along that direction. This point will be further discussed in \cref{sec:equilibrium}.

\subsection{Linearized Dean--Kawasaki equation}
The derivation is so far exact; however, solving \cref{eq:dean} is challenging due to the presence of nonlinearities in both the deterministic and the stochastic term. 
\rev{To make progress, here we linearize it assuming small bath density fluctuations
around a uniform background $\rho_0$~\cite{Demery2014}. 
This strategy is by now quite standard in the physics literature~\cite{Demery2014,Dean:2014,Demery2016,Poncet2017universal,Kruger2017,Kruger2018,Mahdisoltani:2021,venturelli2024universal,muzzeddu2024selfdiffusionanomaliesoddtracer,illien2024deankawasaki}, and has allowed (in spite of the highly singular nature of the Dean--Kawasaki equation~\cite{Cornalba2023}) to derive approximate but accurate predictions for several distinct systems of particles interacting in uniform geometries, under suitable assumptions (discussed below and in~Sec.~\ref{sec:validity}).}
To this end, we consider
\begin{equation}
    \rho(\vb x,t) = \rho_0+\rho_0^{\frac12}\phi(\vb x,t),
    \label{eq:fluctuation}
\end{equation}
and plug it back into \cref{eq:dean}; we then discard small terms\footnote{Note that in fact we are not simply \textit{linearizing} the equation. For instance, the term $\mu \rho_0^{-1}\div \phi(\vb x,t) \grad v(\vb x-\vb r_0)$ turns out to be suppressed by this approximation, although it is linear in $\phi$. However, we adhere here to this commonly adopted terminology.} according to the approximation $\rho_0^{-1/2}\phi \ll 1$. The result reads
\begin{align}
    \dot{\vb r}_0(t)&= 
    -\rho_0^{-\frac12} \hat\mu\z\left( \vb r_0(t)\right)\nabla_{\vb r_0} \int \dd{y} v\left(\vb r_0(t)-\vb{y}\right)\phi(\vb y,t)+\bm\eta_0(\vb r_0(t),t), \label{eq:tracer} \\
    \partial_t \phi(\vb x,t) &= \div \left\lbrace \hat \mu(\vb x)\grad \left[ T  \phi(\vb x,t) + \int\dd{\vb y} u(\vb x-\vb y) \phi(\vb y,t)+  \rho_0^{-\frac12}  v(\vb x-\vb r_0)\right]+\bm \xi (\vb x,t) \right\rbrace, \label{eq:dean_linearized} 
\end{align}
where we rescaled the interaction potentials as
\begin{equation}
    v(\vb x) = \rho_0 U_0(\vb x), \qquad u(\vb x) = \rho_0 U(\vb x).
\end{equation}
Although this approximation becomes formally exact in the dense limit (i.e.~upon sending $\rho_0\to\infty$ by keeping the product $\rho_o \cor V$ fixed, with $\cor V$ being the volume of the system), it has proven to be accurate even away from the dense limit provided that the
interaction potentials $U(\vb x)$ and $U_0(\vb x)$ are sufficiently \textit{soft}, 
so that particles can
overlap completely at a finite energy cost due to thermal fluctuations~\cite{Demery2014,venturelli2024universal}. \rev{This point will be thoroughly discussed in Sec.~\ref{sec:validity}}.

Note that we can still 
\rev{introduce a pseudo free energy}
for the linearized system,
\begin{align}
    \cor F_2[\phi, \vb r_0]\equiv  
    \frac{T}{2}
    \int \dd{\vb x} \phi^2(\vb x) +  \frac 12 \int \dd{\vb x} \dd{\vb y} \phi(\vb x) u(\vb x-\vb y)\phi(\vb y) + \rho_0^{-1/2}  \int \dd{\vb y}\phi(\vb y) v(\vb y-\vb r_0), 
    \label{eq:F2}
\end{align}
so that \cref{eq:dean_linearized} can be rewritten as
\begin{equation}
    \partial_t \phi(\vb x,t) = \div \left[ \hat \mu(\vb x)\grad \fdv{\mathcal F_2}{\phi(\vb x,t)}+\bm \xi (\vb x,t) \right].
\end{equation}
As we did for \cref{eq:reciprocal}, if $v(\vb x)=v(-\vb x)$ we can also rewrite \cref{eq:tracer} as
\begin{equation}
    \dot{\vb r}_0(t)= 
    -\rho_0^{-1/2} \hat\mu\z\left( \vb r_0(t)\right)\nabla_{\vb r_0} \cor F_2[\phi,\vb r_0]+\bm\eta_0(t),
\end{equation}
and similar considerations apply.

\rev{Our derivation is so far quite general, in that the spatial geometry is encoded in the mobility matrices $\hat \mu$ and $\hat \mu\z$ which have not yet been specified. In the following section (and throughout the rest of this paper), we specialize our calculation to the comb geometry.}

\subsection{On the comb}
\label{sec:comb}

Choosing the mobilities as in \cref{eq:mobilities}, the linearized Dean--Kawasaki equation \eqref{eq:tracer} for the tracer particle specializes to
\begin{align}
    \dot{r}_0(t)= 
    -\rho_0^{-1/2} \mu_0\partial_{r_0} \int \dd{\vb y} v\left(\vb r_0(t)-\vb{y}\right)\phi(\vb y,t)+\eta_0(t),
\end{align}
\rev{where $\eta_0$ is a scalar Gaussian noise with zero mean and variance}
\begin{equation}
    \left\langle  \eta_0(t) \eta_0\left(t^{\prime}\right)\right\rangle=2T\mu_0  \delta\left(t-t^{\prime}\right).
\end{equation}
\rev{Indeed, since the tracer can only move along the backbone, its position can be specified by using a single scalar coordinate $r_0(t)$, i.e.~$\vb r_0(t) =  r_0(t)  \vu e_1$}. In the following sections, we will focus first on the evolution of the density 
$\phi$ alone --- namely, we will temporarily switch off the interaction with the tracer in \cref{eq:dean_linearized} \rev{by setting $v(\vb x)=0$,} to obtain
\begin{equation}
    \partial_t \phi(\vb x,t) = \div \left\lbrace \hat \mu(\vb x)\grad \left[ T  \phi(\vb x,t) + \int\dd{\vb y} u(\vb x-\vb y) \phi(\vb y,t)\right]+\bm \xi (\vb x,t) \right\rbrace,
    \label{eq:comb_field_alone}
\end{equation}
where the variance of \rev{the noise} $\bm \xi$ was given in \cref{eq:xi_corr}. This will give access to the dynamical space-time correlations of the density field of interacting bath particles, independently of our initial choice of singling out a tracer particle.
As usual, note that we can rewrite \cref{eq:comb_field_alone} as 
\begin{equation}
    \partial_t \phi(\vb x,t) = \div \left[ \hat \mu(\vb x)\grad \fdv{\cor H}{\phi(\vb x,t)}+\bm \xi (\vb x,t) \right],
    \label{eq:equilibrium_dynamics}
\end{equation}
with
\begin{equation}
    \cor H[\phi]\equiv  \frac{T}{2} \int \dd{\vb x} \phi^2(\vb x) +  \frac 12 \int \dd{\vb x} \dd{\vb y} \phi(\vb x) u(\vb x-\vb y)\phi(\vb y) .
    \label{eq:field_hamiltonian}
\end{equation}
In turn, the dynamical propagator and correlator of $\phi$ \rev{generally} serve as building blocks of the perturbation theory in the presence of a tracer \cite{demerypath,Demery2014,Venturelli_2022,Venturelli_2022_2parts,Venturelli_2023,venturelli2024heat}, whose resulting dynamics will be the subject of \cref{sec:tracer_dynamics}.
\rev{Note that the theory in \cref{eq:equilibrium_dynamics,eq:field_hamiltonian} is quadratic in the density field $\phi$, yet its analysis is made difficult by the presence of the $\delta(\vb x_\perp)$ term in \cref{eq:mobilities}, which manifestly breaks its spatial translational invariance. In the next section, we will show how this difficulty can be overcome by resorting to a self-consistent approach.}

\section{Dynamics of the density fluctuations}
\label{sec:density_dynamics}
The evolution equation~\eqref{eq:comb_field_alone} \rev{for the density field} reads in Fourier space
\begin{equation}
    \partial_t \phi\q(t) =-\mu q_\perp^2\,(T+u\q)\phi\q(t)-\mu q_1^2\,  \int \dperp{p}(T+ u_{q_1,-\vb p_\perp}) \phi_{q_1,\vb p_\perp}(t)+\nu\q(t),
    \label{eq:field_comb}
\end{equation}
where we introduced the Fourier transform $\nu\q(t)$ of the scalar Gaussian noise
\begin{equation}
    \nu(\vb x,t) \equiv \div \bm \xi (\vb x,t),
\end{equation}
with correlations (see \cref{eq:xi_corr})
\begin{equation}
    \expval{\nu\q(t) \nu\p(t')} = 2\mu T\delta(t-t') \left[ q_1^2\delta(q_1+p_1)+q_\perp^2\delta^d(\vb q+\vb p) \right].
    \label{eq:nu_q_corr}
\end{equation}
Here we have used the Fourier convention $f\q(t) = \int \dd[d]{x} \mathrm e^{-\mathrm i\vb{q}\cdot \vb{x}} f(\vb x,t)$, and normalized the Dirac delta in Fourier space as $\int 
\dslash{q}
\delta^d(\vb q)=1$. Below, we will frequently need to adopt mixed representations involving the Fourier or Laplace transform (often with respect to only a few among the spatial and temporal variables on which the observables depend). With a slight abuse of notation, whenever not ambiguous, we will adopt the same symbol (e.g.~$f$) for all these transforms, to avoid the proliferation of hat symbols (such as $\hat f$, $\tilde f$ and so on). As a general convention, we will adopt $\vb x,\vb y,\vb z$ for physical space, $\vb q,\vb p,\vb k$ for Fourier momenta, $t,t'$ for time, and $s,s'$ for Laplace variables.

\rev{To characterize the fluctuations of the particle density $\phi(\vb x,t)$, we now set out to compute its one- and two-point functions. This extends to the case of the comb geometry previous analyses conducted on Brownian particles evolving in homogeneous space~\cite{demerypath,Demery2014,venturelli2024universal} --- see also \ref{app:toy_model}, where some of these expressions are reproduced for convenience.}

\subsection{Propagator}
We start by addressing the propagator, i.e.~the solution of
\begin{equation}
    \partial_t \phi(\vb x,t) = \div  \hat \mu(\vb x)\grad \fdv{\cor H}{\phi(\vb x,t)}+\delta(\vb x,t),
    \label{eq:propagator_def}
\end{equation}
or equivalently in Fourier space
\begin{equation}
    \partial_t  G\q(t) =-\mu q_\perp^2\,(T+u\q) G\q(t)-\mu q_1^2\,
    \int \dperp{p}(T+ u_{q_1,-\vb p_\perp}) G_{q_1,\vb p_\perp}(t) 
    +\delta(t).
    \label{eq:prop-Fourier-def}
\end{equation}
We solve this equation in \ref{app:propagator_comb}; the key step is to realize that, by applying the operator
$\int \dperp{q}(T+u_{q_1,-\vb q_\perp})$ to both members of \cref{eq:prop-Fourier-def}, we can generate a closed self-consistency relation, \rev{which then gives access to the solution}. 
It proves convenient to write the result in the Fourier-Laplace domain, i.e.~$G\q(s)= \int_0^\infty \mathrm \dd{t} \, \mathrm e^{-st} G\q(t)$,
where it reads
\begin{equation}
    G\q(s) = g_\perp(\vb q,s) g_1(q_1,s), 
    \label{eq:comb_prop}
\end{equation}
with
\begin{align}
    &g_\perp^{-1}(\vb q,s) \equiv s+\mu q_\perp^2(T+u\q) , \label{eq:def_gperp} \\
    &g_1^{-1}(q_1,s) \equiv  1+\mu q_1^2 \int \dperp{q} g_\perp(\vb q,s)(T+u_{q_1,-\vb q_\perp}). 
    \label{eq:def_g1}
\end{align}
\rev{As expected,}
by setting $u\q=0$ we correctly recover the moment generating function of a single Brownian particle on the comb, starting at the origin at time $t=0$ --- see \ref{app:propagator_comb}.

Note that, for translationally invariant differential equations, the propagator coincides with the Green's function (i.e.~the solution of \cref{eq:propagator_def} in which $\delta(\vb x) $ is replaced by $\delta(\vb x-\vb x') $), being the latter a function of $\vb x-\vb x'$ only~\cite{Berlioz2024}. By contrast, here the system is only translationally invariant along the backbone $\vb x_\perp=\bm 0$; accordingly, the solution of
\begin{equation}
    \partial_t \phi(\vb x,t) = \div  \hat \mu(\vb x)\grad \fdv{\cor H}{\phi(\vb x,t)}+f(\vb x,t),
    \label{eq:source_def}
\end{equation}
in the presence of a generic source term $f(\vb x,t)$, turns out to read
\begin{equation}
    \phi\q(s) = g_\perp(\vb q,s) f\q(s) -\mu q_1^2 G\q(s) \int \dperp{q} (T+u_{-\vb q}) g_\perp(\vb q,s) f\q(s),
\end{equation}
which only reduces to $\phi\q(s) = G\q(s) f\q(s)$ if $f(\vb x,t)\propto \delta(\vb x_\perp)$.

\subsection{Two-point function: general evolution equation}
\label{sec:twopoint-general}
Computing the two-point function
\begin{equation}
    C\qp(t,t')=\expval{\phi\q(t) \phi\p(t')} 
    \label{eq:comb_correlator_def}
\end{equation}
proves more challenging. To this end, let us first introduce the \textit{response function}
\begin{equation}
    R\qp(t,t') \equiv \expval{\phi\q(t) \nu\p(t')} .
    \label{eq:def_response}
\end{equation}
Upon multiplying the evolution equation~\eqref{eq:field_comb} by $\phi\p(t')$ or $\nu\p(t')$ and taking their average, we obtain the coupled Schwinger-Dyson equations \cite{Itzykson_book,Leticia_1993}
\begin{align}
    \partial_t C\qp(t,t') =& 
    -\mu q_\perp^2\,(T+u\q) C\qp(t,t')-\mu q_1^2\, \int \dperp{k} (T+u_{q_1,-\vb k_\perp})C_{(q_1,\vb k_\perp),\vb p}(t,t')
    \n\\
    &+ R\pq(t',t) ,\label{eq:SD_correlator} \\
    \partial_t R\qp(t,t') =& 
    -\mu q_\perp^2\,(T+u\q) R\qp(t,t')-\mu q_1^2\, \int \dperp{k} (T+u_{q_1,-\vb k_\perp})R_{(q_1,\vb k_\perp),\vb p}(t,t')  \n \\
    &+ 2\mu T \delta(t-t') \left[ q_1^2\delta(q_1+p_1)+q_\perp^2\delta^d(\vb q+\vb p) \right]. 
    \label{eq:SD_response}
\end{align}
We note that, while $C\qp(t,t')$ is enslaved to $R\qp(t,t')$, by contrast \cref{eq:SD_response} 
\rev{provides a closed equation for $R\qp(t,t')$.}
We thus solve first \cref{eq:SD_response} self-consistently in the Laplace domain, as detailed in \ref{app:response_comb}, to obtain
\begin{equation}
    R\qp(s,s') = \frac{2\mu T}{s+s'}g_\perp(\vb p,s) \left[ q_\perp^2\delta^d(\vb q+\vb p)+q_1^2\delta(q_1+p_1)\, s \,G\q(s) \right].
    \label{eq:comb_response_laplace}
\end{equation}
Here and henceforth, we have assumed for simplicity the interaction potential $u(\vb x)$ to be rotationally invariant. The response function in \cref{eq:comb_response_laplace} is symmetric under $(\vb p \leftrightarrow \vb q)$, which was not obvious \textit{a priori}, while the lack of symmetry $(s\leftrightarrow s')$ is expected on causality grounds. Indeed, invoking the properties of the double Laplace transform of causal functions summarized in \ref{app:laplace_causality}, we recognize in \cref{eq:comb_response_laplace} the structure
\begin{equation}
    R\qp(s,s') = \frac{\widetilde R\qp(s)}{s+s'},
    \label{eq:response_causal}
\end{equation}
corresponding in the time domain \rev{(upon taking the inverse Laplace transform with respect to both $s$ and $s'$)} to
\begin{equation}
    R\qp(t,t') = \widetilde R\qp(t-t')\Theta(t-t').
\end{equation}

\subsection{Correlator with Dirichlet initial conditions}
The expression of $R\qp(s,s')$ found in \cref{eq:comb_response_laplace} can now be inserted into \cref{eq:SD_correlator} to obtain a closed equation for the correlator, which we solve self-consistently in \ref{app:correlator_comb} for the case of 
Dirichlet initial conditions $\phi\q(t=0)=\phi\q\z$. 
Recalling that $\phi(\vb x,t)$ represents the fluctuation with respect to a uniform background density \rev{$\rho_0$} (see \cref{eq:fluctuation}), this amounts to imposing quenched initial conditions
\rev{$\rho(\vb x,t=0)=\rho_0 + \rho_0^{1/2} \phi\z(\vb x)$ on the particle density (not necessarily flat).}
The corresponding correlator reads
\begin{align}
    &C\qp(s,s') = \frac{2\mu  T}{s+s'}g_\perp(\vb q,s)g_\perp(\vb p,s') \Bigg\lbrace p_\perp^2\delta^d(\vb q+\vb p) +q_1^2\delta(q_1+p_1)\Bigg[  s' G\q(s')+ s G\p(s) \n\\
    &+  \frac{s'g_1(q_1,s')-sg_1(p_1,s)}{s-s'}\Bigg]\Bigg\rbrace
    +\phi\p\z G\p(s')g_\perp(\vb q,s) \left[ \phi\q\z -\mu q_1^2 g_1(q_1,s)M\z(q_1,s) \right],
    \label{eq:correlator_dirichlet}
\end{align}
with 
\begin{equation}
    M\z(q_1,s) \equiv \int \dperp{q} (T+u_{q_1,-\vb q_\perp})g_\perp(\vb q,s)\phi\q\z.
    \label{eq:def_M0}
\end{equation}
In the case of flat initial conditions, i.e.~$\phi\q\z\equiv\phi\z$, it can be shown that the last term in \cref{eq:correlator_dirichlet} simply reduces to (see \ref{app:correlator_comb})
\begin{equation}
    \left[ \phi\z\right]^2 G\q(s)G\p(s'),
\end{equation}
while the first term represents the connected part of the correlation function,  \rev{i.e.~$\expval{\phi\q(s) \phi\p(s')}_c$}.
Note that the so-obtained correlator is symmetric under the exchange of $(\vb q,s)\leftrightarrow (\vb p,s')$, as expected from its definition in \cref{eq:comb_correlator_def}, but it is not diagonal in the momenta $\vb q$ and $\vb p$ (i.e.~it is not proportional to $\delta^d(\vb q+\vb p)$), because of the spatial anisotropy of the system. However, a common dependence on $\delta(q_1+p_1)$ remains, as a consequence of the translational invariance along the \rev{spatial} direction parallel to the backbone.

\subsection{Equilibrium correlator}
\label{sec:stationary_correlator}

At long times, we expect the two-time correlation function to attain a stationary 
\rev{form}
such that
\begin{equation}
    C\qp (t,t') = \cor C\qp(|t-t'|).
    \label{eq:property_correlator}
\end{equation}
Using the properties of the double Laplace transform summarized in \ref{app:laplace_causality}, this corresponds to
\begin{equation}
    C\qp (s,s') = \frac{\cor C\qp (s) +\cor C\qp (s') }{s+s'},
    \label{eq:corr_laplace_symmetry}
\end{equation}
for some function $\cor C\qp(s)$ that we now set out to compute. 
(By contrast, the Dirichlet correlator given in \cref{eq:correlator_dirichlet} does not have a time-translational invariant structure, \rev{as expected}.)
The simplest way to proceed is to invoke the \textit{fluctuation-dissipation theorem} that relates the correlation function at \textit{equilibrium} to the linear susceptibility
\cite{Tauber}
\begin{equation}
    \chi(\vb x,\vb y,t,t') \equiv  \eval{\fdv{\expval{\phi(\vb x,t)}}{j(\vb y,t')}}_{j=0}.
\end{equation}
The latter is
obtained by adding to the Hamiltonian $\cor H[\phi]$ in \cref{eq:field_hamiltonian} a \rev{source} field $j(\vb x,t)$ linearly coupled to $\phi(\vb x,t)$ as 
\begin{equation}
    \cor H_j[\phi]\equiv  \cor H[\phi]  - \int \dd{\vb x} j(\vb x) \phi(\vb x) ,
    \label{eq:field_hamiltonian_j}
\end{equation}
and by modifying the equation of motion~\eqref{eq:comb_field_alone} \rev{and its solution $\expval{\phi(\vb x,t)}$} accordingly.

As detailed in \ref{app:stationary_correlator_comb}, one can show that the susceptibility \rev{$\chi$} follows an evolution equation very similar to \cref{eq:SD_response} satisfied by the response function \rev{$R$},
and that the two are in fact simply proportional to one another:
\begin{equation}
    R\qp(t,t') = 2T \chi\qp(t,t').
\end{equation}
This is expected, since the noise $\nu\q(t)$ enters linearly in 
\rev{the equation of motion~\eqref{eq:field_comb}.}
Like the response function, the linear susceptibility is also causal and time-translational invariant, that is
\begin{equation}
    \chi\qp(t,t') = \widetilde\chi\qp(t-t')\Theta(t-t').
\end{equation}
By analogy with \cref{eq:response_causal,eq:comb_response_laplace},
we can thus write in the Laplace domain
\begin{equation}
    \widetilde\chi\qp(s) = \mu \, g_\perp(p_\perp,s) \left[ q_\perp^2\delta^d(\vb q+\vb p)+q_1^2\delta(q_1+p_1)\, s \,G\q(s) \right].
\end{equation}
Next, we recall the fluctuation-dissipation theorem~\cite{Tauber}
\begin{equation}
    \dv{\tau} \cor C\qp (\tau) = - T \widetilde\chi\qp(\tau), \qquad \T{for} \; \tau\equiv t-t'>0,
    \label{eq:fdt}
\end{equation}
or equivalently, in the Laplace domain,
\begin{equation}
    \cor C\qp(s) = \frac{1}{s}\left[\cor C\qp(\tau=0^+)-T\chi\qp(s)  \right].
    \label{eq:fdt_laplace}
\end{equation}
The stationary fluctuations $\cor C\qp(\tau=0^+)$ of the field can be accessed by noting that its evolution equation~\eqref{eq:equilibrium_dynamics} admits the equilibrium distribution (see the discussion in \cref{sec:dean})
\begin{equation}
    \cor P[\phi] \propto \mathrm e^{-\frac{1}{T}\cor H[\phi]},
    \label{eq:equilibrium_dist_comb}
\end{equation}
whence (see \ref{app:stationary_correlator_comb}, \rev{and compare e.g.~with Eq.~(43) in Ref.~\cite{demerypath}})
\begin{equation}
    \cor C\qp(\tau=0^+) = \expval{\phi\q(t)\phi\p(t)}_{t\to \infty} = \frac{T}{T+u\q} \delta^d(\vb q+\vb p).
    \label{eq:corr_initial}
\end{equation}
Using \cref{eq:fdt_laplace} then gives, after some straightforward algebra\footnote{
\rev{As a consistency check, one can integrate \cref{eq:equilibrium_correlator_comb} along the backbone by setting $q_1=0$. As expected, one recovers this way the known expression of the two-point function for particles diffusing in homogeneous space --- see the discussion in \ref{app:toy_model}.}
},
\begin{equation}
    \cor C\qp(s) = T g_\perp(\vb p,s) \left[ \frac{\delta^d(\vb q+\vb p)}{T+u\q} -\mu q_1^2\delta(q_1+p_1)G\q(s)  \right].
    \label{eq:equilibrium_correlator_comb}
\end{equation}
This expression admits a closed-form analogous in the time domain, for simple choices of $u\q$ --- see \ref{app:stationary_correlator_comb}.

\begin{figure}
\centering
\includegraphics[]{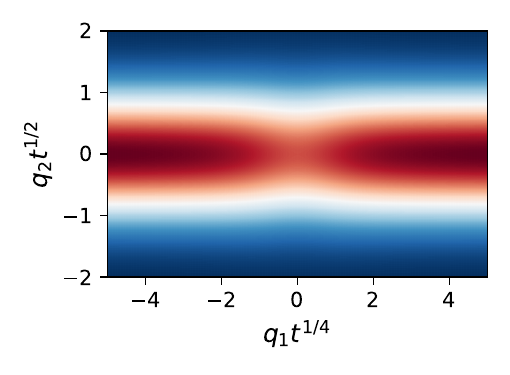}
    \caption{Scaling form assumed at 
    \rev{long}
    times $t$ by the intermediate scattering function $\mathfrak F(\vb q,t)$ --- see \cref{sec:stationary_correlator} and \ref{app:structure} for the details. 
    The amplitude of the ISF is zero in correspondence of the dark-blue regions, and maximum in correspondence of the red regions.
    The plot is obtained for a system of size $\cor{V}=100$ in $d=2$, where the combination $\mu (T+u_{\vb q=\bm 0})$ and the maximum amplitude of the ISF were normalized to unity for illustrative purposes.} 
\label{fig:ISF}
\end{figure}

The stationary correlator in \cref{eq:equilibrium_correlator_comb}, obtained within the linearized SDFT, is one of the key results of this work: it encodes all dynamical \rev{two-point} correlations of the fluctuating particle density. Various dynamical observables, such as structure factors and first passage times, can be accessed starting from \cref{eq:equilibrium_correlator_comb}; as an application, here we consider the \textit{intermediate scattering function} (ISF)
\begin{equation}
    \mathfrak F(\vb q, t) = \frac{1}{N} 
    \expval{\rho\q(t) \rho_{-\vb q}(0)} - \rho_0 \delta^d(\vb q) = \frac{\rho_0}{N}\expval{\phi\q(t) \phi_{-\vb q}(0)} = \frac{\rho_0}{N} \cor C_{\vb q, -\vb q}(t).
\end{equation}
As we demonstrate in \ref{app:structure}, for sufficiently \rev{long} times $t$ the ISF assumes a scaling form $\mathfrak F(q_1 t^\frac14, q_\perp t^\frac12)$
that is independent of the specifics of the interaction potential $u(\vb x)$, and which we compute explicitly for the case $d=2$ (see \cref{eq:scaling-ISF}). The resulting scaling function is plotted in \cref{fig:ISF} for selected choices of the various parameters, and with its amplitude rescaled to unity for graphical convenience. Such ISF is manifestly reminiscent of the anisotropy of the comb structure.

\subsection{Proof that the stationary and equilibrium correlators coincide}
\label{sec:equilibrium}
In the previous section we computed the equilibrium correlator, see \cref{eq:equilibrium_correlator_comb}.
However, as we stressed under \cref{eq:stationary_dist},
there is no reason to believe \textit{a priori} that the stationary state of the system is also an \textit{equilibrium} one. To this end, here we prove that the solution obtained in \cref{sec:stationary_correlator}
is unique --- namely, that no other function exists for this system with the property of time-translational invariance required in \cref{eq:property_correlator}.

We start by noting that an alternative and more direct way to impose the time-translational invariance $C\qp(t,t')=\cor C\qp(t-t')$ of the correlator (without invoking the fluctuation-dissipation theorem, which holds only at equilibrium) is to require
\begin{equation}
    (\partial_t+\partial_{t'})C\qp(t,t')=0.
    \label{appeq:station_condition}
\end{equation}
This condition can be used to construct the stationary correlator as follows. First, we change indices in the field equation~\eqref{eq:field_comb} and write an evolution equation for $\partial_{t'}\phi\p(t')$. Next, we multiply it by $\phi\q(t)$ and take its average over the noise. This gives another Schwinger-Dyson equation,
\begin{align}
    \partial_{t'} C\qp(t,t') =& 
    -\mu p_\perp^2\,(T+u\p) C\qp(t,t')-\mu p_1^2\, \int \dperp{k} (T+u_{p_1,-\vb k_\perp})C_{\vb q,(p_1,\vb k_\perp)}(t,t')
    \n\\&+ R\qp(t,t'),
\end{align}
akin to \cref{eq:SD_correlator}, which can then be summed to the latter. Setting the result equal to zero as per \cref{appeq:station_condition}, 
and upon introducing
\begin{align}
    &\cor G\qp^{-1} \equiv \mu \left[ q_\perp^2\,(T+u\q)+p_\perp^2\,(T+u\p) \right], \label{appeq:def_Gqp}\\
    &\cor S\qp(t,t') \equiv R\qp(t,t')+R\pq(t',t), 
    \label{appeq:source}
\end{align}
we find \rev{that the stationary correlator $ C\qp(t,t')$ must satisfy}
\begin{align}
    C\qp(t,t') = & \,
    \cor G\qp\, \cor S\qp(t,t')  -\mu \cor G\qp \Bigg[  q_1^2\, \int \dperp{k} (T+u_{q_1,-\vb k_\perp})C_{(q_1,\vb k_\perp),\vb p}(t,t')\n\\
    &+ p_1^2\, \int \dperp{k} (T+u_{p_1,-\vb k_\perp})C_{\vb q,(p_1,\vb k_\perp)}(t,t') \Bigg].
    \label{appeq:double_SD}
\end{align}
This integral equation is analyzed in \ref{app:uniqueness}; 
\rev{there, we demonstrate that in fact \cref{appeq:double_SD}} 
admits a unique solution, corresponding to the equilibrium correlator found in \cref{eq:equilibrium_correlator_comb}. \rev{This concludes our analysis of the one- and two-point functions within the linearized Dean--Kawasaki theory on the comb, which completely characterize (within this Gaussian approximation) the particle density fluctuations.}

\section{Tracer particle dynamics}
\label{sec:tracer_dynamics}

\rev{In \cref{sec:density_dynamics} we have analyzed the fluctuations of the bath density, assuming that all bath particles are equal. In this Section, instead, we go back to \cref{eq:dean_linearized,eq:tracer}, where a tagged tracer (singled out from the particles bath) evolves in contact with the fluctuating density of surrounding particles.}
The resulting coupled equations for the tracer and the bath density read, in Fourier space,
\begin{equation}
    \dot{r}_0(t)= 
    -h\mu_0  \int \dslash{q} i q_1 v_{\vb q}\phi_{\vb q}(t)\mathrm e^{i q_1 r_0(t)}+\eta_0(t), \label{eq:tracer_fourier_comb} 
\end{equation}
with $\expval{\eta_0(t)\eta_0(t')}= 2\mu_0 T\delta(t-t' )$,
and
\begin{align}
     \partial_t \phi\q(t) =&\,
     -\mu q_\perp^2\,(T+u\q)\phi\q(t)-\mu q_1^2\,  \int \dperp{p}(T+ u_{q_1,-\vb p_\perp}) \phi_{q_1,\vb p_\perp}(t) +\nu\q(t)\n\\
     &-h\mu \left[ q_\perp^2v\q+ q_1^2 v_{q_1}(\vb x_\perp =\bm 0)  \right]\mathrm e^{-i q_1 r_0(t)},
     \label{eq:field_fourier_comb}
\end{align}
with the correlations of $\nu\q(t)$ given in \cref{eq:nu_q_corr}. Above we denoted for brevity $h\equiv \rho_0^{-1/2}$.

With this in hands, in the next two sections we tackle two distinct (but related) problems using perturbation theory for small $h$:
\begin{enumerate}[(i)]
    \item In \cref{sec:general_prof_constrained} we analyze the generalized correlation profiles between the tracer position and the density of surrounding bath particles~\cite{venturelli2024universal}. \rev{As explained below, these quantities encode the response of the bath 
    to the presence of a moving inclusion.}
    \item In \cref{sec:MGF} we use such correlation profiles to access the effective tracer particle dynamics, and in particular its effective diffusion coefficient~\cite{Demery2014}.
\end{enumerate}
Note that, being constrained to move only along the backbone, \rev{we expect that} the tracer particle will 
always exhibit diffusive behavior in spite of the comb geometry (whereas the other bath particles sub-diffuse along the backbone, \rev{as recalled in \cref{sec:model}}).

\subsection{Generalized correlation profiles}
\label{sec:general_prof_constrained}
The simplest measure of the cross correlations between the tracer position and the density of surrounding bath particles is the average density profile in the reference frame of the tracer:
\begin{equation}
    \psi(\vb x,t) \equiv \expval{\phi(\vb x +\vb r_0(t),t)  } \qquad \to \qquad \psi\q(t) = \expval{\phi_{\vb q}(t)\mathrm e^{i \vb q \cdot \vb r_0(t)} } ,
    \label{eq:hq_def}
\end{equation}
where $\vb r_0(t) = r_0(t)\, \vu e_1$.
\rev{This quantity has been first analyzed in Ref.~\cite{Demery2014} for the case of interacting Brownian particles evolving in uniform space.
In particular,
in the presence of an external bias applied to the tracer (which drives the system out of equilibrium), the average stationary density $\psi(\vb x)$ has been proven in~\cite{Demery2014} to exhibit an \textit{algebraic tail} in the direction opposite to that of the bias. This long-range effect is in fact a signature of the many-body interacting nature of the system. In the absence of a bias (i.e.~at equilibrium), $\psi(\vb x)$ was instead shown to be a rapidly decaying function.}

\rev{For the case of particles interacting on a comb (with the tagged tracer constrained along the backbone), the average profile $\psi(\vb x,t)$}
can be evaluated by using Stratonovich calculus and the Novikov theorem, as delineated in \ref{app:constrained_profiles}.
\rev{In the stationary state attained by the system at long times, the result reads in Fourier space}
\begin{align}
    \psi\q &= -h\frac{v\q }{T+u\q} +\order{h^2}. \label{eq:hq}
\end{align}
Crucially, 
this
prediction is exactly the same as
\rev{that derived in~\cite{Demery2014} for particles evolving in}
homogeneous $d$-dimensional space, i.e.~without the comb constraint~\cite{venturelli2024universal}.
\rev{Although this may sound surprising, in fact} 
a non-perturbative calculation\footnote{See Sec.~II.D.4 in the Supplemental Material of Ref.~\cite{venturelli2024universal}.} 
\rev{reveals that \cref{eq:hq} follows}
directly from the equilibrium distribution in \cref{eq:stationary_dist}, showing once again that the latter correctly represents the stationary state of the system.

However, signatures of the comb structure are expected to show up 
by considering higher-order correlation profiles such as
\begin{equation}
    g(\vb x,t) \equiv \expval{r_0(t) \phi(\vb x +\vb r_0(t),t)  } \qquad \to \qquad  g_{\vb q}(t) = \expval{ r_0(t) \phi_{\vb q}(t)\mathrm e^{i \vb q \cdot \vb r_0(t)} } .
    \label{eq:gq_def}
\end{equation}
\rev{The information encoded in $g(\vb x,t)$ is the following: how does the bath density at distance $\vb x$ from the tracer fluctuate, in response to a fluctuation of the tracer position $\vb r_0(t)$? Similarly,}
all correlations profiles of the form $\expval{[r_0(t)]^n \phi(\vb x +\vb r_0(t),t)  }$ (for integer $n$) are compactly encoded in the generating function~\cite{Grabsch_2023}
\begin{equation}
    w(\vb x,\lambda,t) \equiv \frac{\expval{  \phi(\vb x +\vb r_0(t),t)\,\mathrm e^{ \lambda \, r_0(t)}}}{ \expval{\mathrm e^{ \lambda \, r_0(t)}} }  \qquad \to \qquad w_{\vb q}(\lambda, t) = \frac{\expval{  \phi_{\vb q}(t) \mathrm e^{ (\lambda +i q_1)  r_0(t)}}}{ \expval{\mathrm e^{ \lambda \, r_0(t)}} } ,
    \label{eq:wq_def}
\end{equation}
from which they can be retrieved by taking derivatives with respect to $\lambda$, before setting $\lambda =  0$.

\rev{An analysis of these profiles for particles interacting in homogeneous space has been recently presented in Ref.~\cite{venturelli2024universal}, with a special focus on the stationary profile $g(\vb x)$.
The main finding was that $g(\vb x)$ exhibits a nontrivial large-distance behavior also in the absence of a bias: in fact, it displays a long-range power-law decay $g(x)\sim x^{1-d}$, whose exponent does not depend on the details of the interaction potential $u(\vb x)$, but actually only on the spatial dimensionality $d>1$ of the system. This surprising result has been checked to hold for a variety of systems (including hard-core lattice gases and Lennard--Jones suspensions), using both analytical calculations and numerical simulations.}

\rev{On the comb, these generalized correlation profiles can again}
be evaluated by using Stratonovich calculus and the Novikov theorem, as shown in \ref{app:constrained_profiles}. In particular, 
\rev{their
generating function~\eqref{eq:wq_def}}
reads in the stationary limit\footnote{\rev{Its counterpart in homogeneous space is reported for comparison in \cref{eq:wq_sol}. Similarly, the profile $g\q$ in \cref{eq:gq_nof} should be compared with Eq.~(12) in Ref.~\cite{venturelli2024universal}, valid in homogeneous space.}}
\begin{align}
        w\q(\lambda) = h G_0(\vb q,\lambda) \Bigg\lbrace &
        \frac{v\q}{T+u\q}\left[i \mu_0 T q_1 (\lambda +i q_1) - \mu q_\perp^2(T+u\q)  \right] \n\\
        &-\mu q_1^2 \frac{v_{q_1}(\vb x_\perp = \bm 0) +M\f(q_1,\lambda)}{1+\mu q_1^2 M\tr(q_1,\lambda)} 
        \Bigg\rbrace +\order{h^2},
        \label{eq:wq}
\end{align}
where \rev{we introduced}
\begin{align}
    G_0^{-1}(\vb q,\lambda) &\equiv \mu q_\perp^2(T+u\q)-i\mu_0 T q_1 (2\lambda+i q_1) 
    , \label{eq:G0}\\
    M\tr(q_1,\lambda) &\equiv \int \dperp{q} (T+u_{-\vb q} )\, G_0(\vb q,\lambda),\\
    M\f(q_1,\lambda) &\equiv  \int \dperp{q} v\q\, G_0(\vb q,\lambda) \left[ i\mu_0 q_1 T(\lambda+iq_1)-\mu q_\perp^2(T+u_{\vb q} )  \right].
\end{align}
Similarly, 
the correlation profile $g\q(t)$ introduced in \cref{eq:gq_def} turns out to reduce in the stationary state to
(see \ref{app:constrained_profiles})
\begin{align}
    g\q &= h 
    \frac{i\mu_0 Tq_1}{\mu q_\perp^2(T+u\q)+\mu_0 T q_1^2}
    \left[ \frac{
    q_1^2 M\t(q_1)}{1+
    q_1^2 M\o(q_1)}-\frac{v\q}{T+u\q} \right]  +\order{h^2},
    \label{eq:gq_nof}
\end{align}
where we called
\begin{align}
    M\o(q_1) &\equiv \int \dperp{q} \frac{
    \mu
    (T+u_{-\vb q} )}{\mu q_\perp^2(T+u\q)+\mu_0 T q_1^2}, \label{eq:M1}\\
    M\t(q_1) &\equiv  \int \dperp{q} \frac{
    \mu
    v\q}{\mu q_\perp^2(T+u\q)+\mu_0 T q_1^2}
    .
    \label{eq:M2}
\end{align}
By symmetry in the momenta $\vb q$, we deduce that $g(- \vb x)=-g(\vb x)$, as expected from its definition~\eqref{eq:gq_def} --- see also the inset of \cref{fig:tracer}(a).
Moreover, if $\mu_0=\mu$, then the stationary correlation profile $g\q$ becomes independent of the particles mobility.
In $d=2$ and setting $v\q=u\q$, it is then simple to derive its large-distance behavior (see \ref{app:constrained_profiles})
\begin{equation}
    g(x_1) \sim \frac{h \sqrt{T} u_{\vb q=\bm 0}}{2\pi (T+u_{\vb q=\bm 0})^{3/2}}x_1^{-1}.
    \label{eq:g-large-distance}
\end{equation}
This asymptotic behavior is checked in \cref{fig:tracer}(a).
Remarkably, it is consistent with the behavior $g(x_1) \sim x_1^{1-d}$ observed in uniform $d$-dimensional space (i.e.~without the comb) in Ref.~\cite{venturelli2024universal}.
\rev{In this sense, \cref{eq:g-large-distance} further validates the claim of universality of the large-distance behavior of the correlation profile $g(\vb x)$, whose decay exponent was found in~\cite{venturelli2024universal} to depend only on the spatial dimensionality $d$ of the system.}

\subsection{Effective diffusion coefficient}
\label{sec:MGF}

In this Section we focus on the effective dynamics of the tracer particle, and we derive the first perturbative correction to its diffusion coefficient, due to the presence of the other bath particles. First, note that the leading-order correction to the generalized profiles computed in \cref{sec:general_prof_constrained} turned out to be of $\order{h}$. Conversely, any correlation function involving $\vb r$ but not $\phi$ must exhibit corrections at least of $\order{h^2}$ --- the simplest way to prove this fact is to note that the system of equations~\eqref{eq:tracer_fourier_comb}--\eqref{eq:field_fourier_comb} is invariant under the transformation $\{h\mapsto -h,\phi \mapsto -\phi\}$~\cite{wellGauss,Venturelli_2022}.
In general, the perturbative calculation of the effective diffusion coefficient in homogeneous space can be compactly addressed within the path-integral formalism developed in Ref.~\cite{demerypath} (and later adopted in e.g.~Refs.~\cite{Demery2014,Benois2023}). 
However, it is not evident how to extend such path-integral techniques to the case of non-homogeneous space analyzed here.
By contrast, we showed in Ref.~\cite{venturelli2024universal} that the tracer-bath correlation profiles --- that we already derived in the previous Section --- actually encode the tracer statistics, and thus offer a straightforward alternative way to access its diffusion coefficient.

\begin{figure}
\centering
\includegraphics[width=0.45\textwidth]{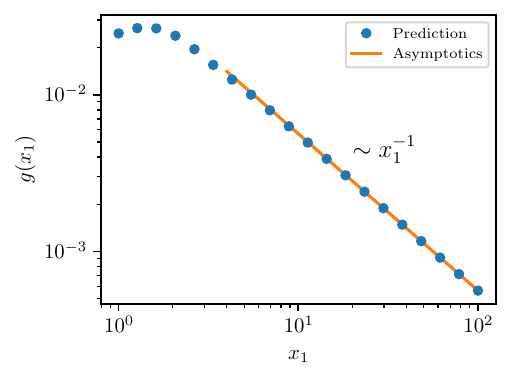}
  \put(-197,135){(a)}
\put(-160,30){\includegraphics[scale=0.325]{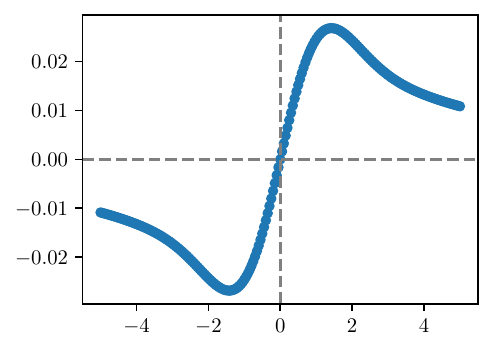}}
  \hspace{10pt}
\includegraphics[width=0.45\textwidth]{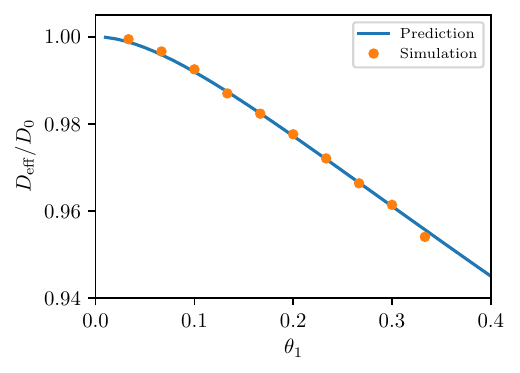}
  \put(-200,135){(b)}
    \caption{\textbf{(a)} Tracer-bath spatial correlation profile $g(x_1)$, whose Fourier transform is predicted by \cref{eq:gq_nof} in the long-time stationary state. Its asymptotic scale-free behavior is given in \cref{eq:g-large-distance}, while the inset shows its symmetry $g(-\vb x)=-g(\vb x)$. In the plot we chose a Gaussian interaction potential $u\q=\exp(-q^2/2)$, and we set $h$ and $T$ to unity for simplicity. \textbf{(b)} Comparison between the effective diffusion coefficient $D_\text{eff}$ predicted by \cref{eq:effective_diff}, and the results of numerical simulations 
    of a system of particles interacting via Gaussian potentials of strength $\epsilon$.
    In the plot, we vary $\epsilon$ (and hence the dimensionless parameter $\theta_1\equiv \epsilon \rho_0 \sigma^d/ T$, see \cref{sec:MGF}), with all other parameters kept fixed. Details of the numerical simulation are reported in \ref{app:simulation}.} 
\label{fig:tracer}
\end{figure}

To show this, let us focus on the moment generating function
\begin{equation}
    \varphi(q_1,t) \equiv \expval{\mathrm e^{-iq_1  r_0(t)}},
    \label{eq:mgf_const}
\end{equation}
that we aim to compute
in the stationary regime. To this end, one can first use Stratonovich calculus to derive,
starting from \cref{eq:tracer_fourier_comb,eq:field_fourier_comb},
the relation\footnote{See Sec.~II.G in the Supplemental Material of Ref.~\cite{venturelli2024universal}.}
\begin{equation}
    \partial_t \expval{\mathrm e^{\lambda \, r_0(t)}} = \lambda \expval{\eta (t) \mathrm e^{\lambda \, r_0(t)}} -h\mu_0  \lambda \cdot \int \dslash{p} i  p_1 v\p \expval{\mathrm e^{( \lambda+i p_1)  r_0(t)} \phi\p(t) }.
    \label{eq:rel_strat}
\end{equation}
The second expectation value on the r.h.s.~of \cref{eq:rel_strat} is 
reminiscent of
a quantity we have already met, \rev{i.e.~the generating function~$w\p(\lambda,t)$ of tracer-bath correlation profiles} introduced in \cref{eq:wq_def}. Since the evolution equation \rev{of} $w\p(\lambda,t)$ (derived perturbatively in \ref{app:wq}) admits as a stationary solution $w\p(\lambda)$ given in \cref{eq:wq_def},
here it is reasonable to assume that
\begin{equation}
    \expval{\mathrm e^{( \lambda+i p_1)  r_0(t)} \phi\p(t) } = w\p(\lambda,t) \expval{\mathrm e^{\lambda \, r_0(t)}} \simeq w\p(\lambda) \expval{\mathrm e^{\lambda \, r_0(t)}},
\end{equation}
in the stationary regime.
Plugging this into \cref{eq:rel_strat} then gives
\begin{equation}
    \partial_t \expval{\mathrm e^{\lambda \, r_0(t)}} =  \lambda \expval{\eta (t) \mathrm e^{\lambda \, r_0(t)}} -h\mu_0 \expval{\mathrm e^{\lambda \, r_0(t)}}  \lambda_1  \int \dslash{p} i  p_1 v\p  w\p(\lambda).
    \label{eq:constr_partial}
\end{equation}
On the other hand, the first expectation value on the r.h.s.~of \cref{eq:rel_strat,eq:constr_partial} can be computed by standard methods using the Novikov theorem~\cite{Luczka_2005}, and the result reads
\begin{equation}
    \expval{\eta (t) \mathrm e^{ \lambda \, r_0(t)}} = \mu_0 T \lambda \expval{ \mathrm e^{ \lambda \, r_0(t)}}, 
\end{equation}
which is the same as for a non-interacting tracer (in spite of the interaction with the bath density, see Ref.~\cite{venturelli2024universal} for further details). 
By setting $\lambda=-iq_1$ and using \cref{eq:mgf_const},
we thus obtain
\begin{align}
    \label{eq:dt_varphi}
    \partial_t \varphi(q_1,t) = -q_1^2\mu_0 T \varphi(q_1,t) -h \mu_0 \varphi(q_1,t) \int \dslash{p} (p_1\cdot q_1) v\p  w\p(-iq_1),
\end{align}
whose solution\footnote{Note that \cref{eq:logPsi} apparently implies that \rev{not only the variance, but actually} all other higher-order cumulants of $r_0(t)$ also grow linearly with $t$. However, we stress that \cref{eq:logPsi} has been derived under the assumption that $w\q(\lambda,t)$ becomes stationary, and must thus be taken with caution --- we comment further on this point in \ref{app:uniform_kurtosis}.}
\begin{align}
    \ln \varphi(q_1,t) =-\mu_0 t \left[ q_1^2 T  +h  \int \dslash{p} (p_1\cdot q_1) v\p w\p(-iq_1)\right] 
    \label{eq:logPsi}
\end{align}
gives the leading-order correction to the cumulant generating function of the tracer position.
\rev{Thus, as anticipated above, \cref{eq:logPsi} links the statistical properties of the tracer position $r_0(t)$ to the knowledge of the tracer-bath correlation profiles encoded in $w\p(\lambda)$. In fact, this expression}
can be made explicit by using the leading-order estimate of $w\p(\lambda)$ found in \cref{eq:wq}. \rev{Several comments are now in order.}
First, we note that \rev{the tracer position $ r_0(t)$} has non-Gaussian statistics \rev{(indeed, the r.h.s.~of \cref{eq:logPsi} is not simply quadratic in $q_1$)}.
In particular, one can check that $\varphi(q_1= 0,t)=1$ and \rev{that the average tracer position vanishes,}
\begin{equation}
    \expval{r_0(t) } = \eval{i\frac{\partial \varphi(q_1,t)}{\partial q_1}}_{q_1=0} = 0,
\end{equation}
as expected \rev{by the symmetry of the problem under $x_1 \mapsto - x_1$ (indeed, we have assumed that the tracer starts its motion at the origin, i.e.~$r_0(t=0)=0$). Conversely, its variance reads}
\begin{align}
    \expval{ r_0^2(t)  } &= \eval{-\frac{\partial^2}{\partial q_1^2 }\varphi(q_1,t)}_{q_1= 0} 
    \label{eq:variance}
    \\
    &= 2\mu_0 T t \left\lbrace 1- h^2 \int \frac{\dd{p_1}}{2\pi} p_1^2 \left[M^{(5)}(p_1)  
    -
    \frac{\mu_0}{\mu} 
      \frac{ [p_1 M\t(p_1)]^2}{1+
    p_1^2 M\o(p_1)} 
    \right]\right\rbrace+\order{h^4},\n
\end{align}
where 
\begin{equation}
    M^{(5)}(p_1) \equiv \int \dperp{p} 
    \frac{
    \mu_0
    v\p^2 }
    {(T+u_{\vb p})[\mu q_\perp^2(T+u\q)+\mu_0 T q_1^2] },
\end{equation}
while the functions $M^{(1,2)}(p_1)$ were given in
\cref{eq:M1,eq:M2}, respectively.
Calling $D_0=\mu T$ the bare diffusion coefficient of the non-interacting tracer, from \cref{eq:variance} we thus obtain
$\expval{r_0^2(t)}=2 D_\T{eff} t$ with the effective diffusion coefficient
\begin{equation}
    \frac{D_\T{eff}}{D_0} = 1- h^2 \int \frac{\dd{p_1}}{2\pi} p_1^2 \left[M^{(5)}(p_1)  
    -
    \frac{\mu_0}{\mu} 
      \frac{ [p_1 M\t(p_1)]^2}{1+
    p_1^2 M\o(p_1)} 
    \right]+\order{h^4}.
    \label{eq:effective_diff}
\end{equation}
This prediction represents another key result of this work, and is the spatially heterogeneous counterpart to the diffusion coefficient first obtained \rev{in Ref.~\cite{Demery2014} for homogeneous media, using path-integral methods (see Eq.~(52) therein)}.

Again, we note that for $\mu_0=\mu$ 
the second term in 
\cref{eq:effective_diff} actually
becomes $\mu$-independent.
To be concrete,  we now set $v\q=u\q$, and assume for the interaction potential in real space $U(\vb x)$ the general rotationally invariant form
\begin{equation}
    U(\vb x) = \epsilon f(x/\sigma),
    \label{eq:pot1}
\end{equation}
where $ x=\abs{\vb x}$, $\epsilon$ is an energy scale, $\sigma$ is a microscopic length scale, and the function $f$ is dimensionless. For example, within the Gaussian core model~\cite{Lang2000} one has $U(x)=\epsilon \exp[-x^2/(2\sigma^2)]$. In Fourier space, the corresponding rescaled interaction potential $u\q = \rho_0 U\q $ then reads
\begin{equation}
    u\q = \rho_0\epsilon \, \sigma^d f(q\sigma).
     \label{eq:pot2}
\end{equation}
By inspecting \cref{eq:effective_diff} with $\mu_0=\mu$,
and upon reinstating the microscopic length scale $a$ by replacing $q_1^2 \mapsto a^{d-1} q_1^2$ (see the discussion 
in \cref{sec:model}),
it is then simple to see that $D_\T{eff}/D_0$
only depends on the 
dimensionless
combinations $\theta_1\equiv \epsilon \rho_0 \sigma^d/ T$, $\theta_2=\rho_0\sigma^d$, and $\theta_3=R/a$, rather than on these parameters taken separately.
In \cref{fig:tracer}(b) we thus compare our prediction in~\cref{eq:effective_diff} to the results of numerical simulations obtained by varying the interaction strength $\epsilon$ with all other parameters fixed (see \ref{app:simulation} for the details), which amounts to varying the effective parameter $\theta_1$. The prediction is shown to be accurate for small interaction strengths $\epsilon \ll k_B T$, as usually expected within the linearized Dean--Kawasaki theory~\cite{Demery2014,venturelli2024universal} \rev{--- this point is further discussed in the next section}.

\subsection{Validity of the linearized theory}
\label{sec:validity}

\rev{We conclude our discussion by briefly recalling the assumptions under which our calculation is expected to accurately reproduce the results of numerical simulations. These assumptions are not specific to the comb geometry, but actually underlie similar treatments (even in uniform space) based on the linearized Dean--Kawasaki theory~\cite{Demery2014}:
(i)~The system should admit a well-defined average background density $\rho_0$ (see \cref{eq:fluctuation}); note that, for very dilute systems with attractive potentials, clustering phenomena may take place and undermine this assumption; (ii)~The interaction potentials should admit a Fourier transform, which excludes strong short-distance repulsion (such as the one provided by Lennard--Jones-type potentials); (iii)~The system should not be too dilute, our results being expressed in a power series in  the small parameter $h=\rho_0^{-1/2}$.
More precisely, the latter parameter \textit{not} being dimensionless, an estimate of the limits of validity of the perturbative expansion can be rather formulated as~\cite{venturelli2024universal}
\begin{equation}
    \frac{\epsilon}{k_B T} \ll \sigma^{-d} \left[\rho_0 \int \dslash{q} f^2(q\sigma)  \right]^{-1/2},
    \label{eq:criterium-2}
\end{equation}
depending on the specific details of the interaction potential~\eqref{eq:pot1}. This effectively conveys that the linearized Dean--Kawasaki theory is an expansion around the weakly interacting case, $\epsilon$ being the typical strength of these interactions.}

\section{Conclusions}

In this work we considered a system of overdamped Brownian particles interacting via soft pairwise potentials, and constrained to move on a comb-like structure (see \cref{fig:comb}).
Within the Dean--Kawasaki formalism, we first derived in \cref{sec:dean} the exact 
equations that describe the dynamics of the particle density field $\rho(\vb x,t)$. The latter can be
expanded around a uniform background density, 
which allowed us to characterize the Gaussian fluctuations of the density field. In particular, in \cref{sec:density_dynamics} we computed the density-density correlator both starting from fixed initial conditions (see \cref{eq:correlator_dirichlet}), and in the stationary limit assumed by the system at long times (see \cref{eq:equilibrium_correlator_comb}). The latter can be used to construct for instance the intermediate scattering function shown in \cref{fig:ISF}.
To the best of our knowledge, this represents the first application of the Dean--Kawasaki theory to non-homogeneous media~\cite{teVrugt2020,illien2024deankawasaki}.

Furthermore, in \cref{sec:tracer_dynamics} we singled out a tagged tracer from the bath of interacting particles, under the assumption that it is constrained to move only along the backbone.
In this setting, we derived the spatial correlation profile between the tracer position and the surrounding bath density (see \cref{eq:gq_nof} and \cref{fig:tracer}(a)), and the effective diffusion coefficient of the tracer, which we tested using Brownian dynamics simulations (see \cref{eq:effective_diff} and \cref{fig:tracer}(b)).

Further extensions of this work could address the \textit{unconstrained} problem in which the tracer particle is allowed to leave the backbone (which is technically more challenging than the constrained case considered in this work, even within the perturbation theory delineated in \cref{sec:tracer_dynamics}). This would ideally provide access to the effective \textit{subdiffusion} coefficient of the tracer along the backbone, i.e.~to its correction due to the presence of the other bath particles. 
Moreover, the framework delineated in \cref{sec:dean} can in principle be extended to other non-homogeneous geometries \rev{(such as  fractal
networks and random percolation structures)}, which can be encoded with a suitable choice of the particle's mobility matrix $\hat \mu(\vb x)$ in \cref{eq:langevin}. 

\ack
DV thanks Leticia F. Cugliandolo, Pietro Luigi Muzzeddu, and Gilles Tarjus for illuminating discussions.

\appendix
\addtocontents{toc}{\fixappendix}

\section{Details on the dynamics of the density fluctuations}
\label{app:density}
In this Appendix we completely characterize the density fluctuations on the comb by deriving the results reported in \cref{sec:density_dynamics}. In particular, in \ref{app:propagator_comb}, \ref{app:response_comb}, and \ref{app:correlator_comb} we provide the derivations of the propagator, response function, and Dirichlet correlator, respectively. In \ref{app:stationary_correlator_comb} we then derive the stationary correlator, proving in \ref{app:uniqueness} that it coincides with the equilibrium correlator (in spite of the geometrical constraints of the system, which could in principle preclude ergodicity). In \ref{app:laplace_causality} we recall a few causality properties of the double Laplace transform, and finally in \ref{app:toy_model} we analyze in the Laplace domain the free (Gaussian) field~\cite{Tauber}, as a test bench for the techniques employed on the comb.

\subsection{Derivation of the propagator}
\label{app:propagator_comb}

We start from \cref{eq:prop-Fourier-def}, which rules the evolution of the Fourier transform $G\q(t)$ of the propagator $G(\vb x,t)$.
We then Laplace-transform the last equation:
using that $G(\vb x,t=0^+)=0$ by construction, we find
\begin{equation}
    G\q(s) = g_\perp(\vb q_\perp,s)\left[ 1-\mu q_1^2  F (q_1,s) \right],
    \label{appeq:transformed_prop}
\end{equation}
where $g_\perp$ was defined in \cref{eq:def_gperp}, while we introduced the auxiliary function
\begin{align}
     F (q_1,s) = \int \dperp{q}(T+u_{q_1,-\vb q_\perp}) G\q(s).
\end{align}
Crucially, the terms enclosed in square brackets in \cref{appeq:transformed_prop} do not depend on $\vb q_\perp$. Upon applying 
$\int \dperp{q}(T+u_{q_1,-\vb q_\perp})$ to \cref{appeq:transformed_prop}, we thus obtain the
self-consistent equation
\begin{equation}
    F (q_1,s) = M(q_1,s)\left[ 1-\mu q_1^2  F (q_1,s) \right],
    \label{appeq:self_prop}
\end{equation}
where we introduced the function
\begin{align}
     M (q_1,s) = \int \dperp{q}(T+u_{q_1,-\vb q_\perp}) g_\perp(\vb q_\perp,s).
     \label{appeq:def_M}
\end{align}
We thus solve \cref{appeq:self_prop}
to obtain $F$, which can then be replaced into \cref{appeq:transformed_prop} to give
\begin{equation}
    G\q(s) = \frac{g_\perp(\vb q_\perp,s)}{1+\mu q_1^2 M(q_1,s)}.
    \label{appeq:prop_second_last}
\end{equation}
Upon calling (compare with \cref{eq:def_g1})
\begin{equation}
    g_1^{-1}(q_1,s) \equiv 1+\mu q_1^2  M (q_1,s)
\end{equation}
the denominator in \cref{appeq:prop_second_last},
we obtain the expression of $G\q(s)$ reported in \cref{eq:comb_prop}.

Note that \cref{eq:prop-Fourier-def} reduces, for $u\q=0$ and in $d=2$, to the Fourier transform of the Fokker-Planck equation~\eqref{eq:single_comb} for a single (non-interacting) Brownian particle on the comb, placed initially at the origin at time $t=0$. 
The corresponding solution $G\q(s)$, i.e.~its moment generating function written in the Laplace domain, then reduces to
\begin{equation}
    G\q(s) = (s+\mu T q_2^2)^{-1} \left[  1+\frac{a\sqrt{\mu T} q_1^2}{2\sqrt{s}}\right]^{-1},
    \label{eq:MGF-free-part}
\end{equation}
where we computed explicitly the integral in the definition~\eqref{eq:def_g1} of $g_1(q_1,s)$, and we reinstated the length scale $q_1^2\mapsto a q_1^2$ (see the discussion 
in \cref{sec:model}).
In particular, one can use the Tauberian theorems~\cite{hull1955asymptotic} to deduce the long-$t$ behavior of the inverse Laplace transform of the particle variance along the two orthogonal directions:
\begin{equation}
    -\eval{\pdv[2]{G\q(s)}{q_2}}_{\vb q=\bm 0} = 2\mu T s^{-2} \qquad \leftrightarrow \qquad \expval{y^2(t)}\simeq 2\mu T t,
\end{equation}
and
\begin{equation}
    -\eval{\pdv[2]{G\q(s)}{q_1}}_{\vb q=\bm 0} = a \sqrt{\mu T} s^{-\frac32} \qquad \leftrightarrow \qquad \expval{x^2(t)} \simeq 2a \sqrt{\frac{\mu T t}{\pi} } .
\end{equation}
Alternatively, one can invert explicitly \cref{eq:MGF-free-part} to the time domain, finding
\begin{equation}
    G\q(t) =\frac{4q_2^2 \mathrm e^{-q_2^2   \mu T t}  - 4 a q_1^2 \sqrt{\frac{q_2^2}{\pi}} \,\mathrm{DawsonF}\left(\sqrt{q_2^2 \mu T t}\right) + a^2 \mathrm e^{\frac{q_1^4}{4} a^2  \mu T t} q_1^4 \,\mathrm{Erfc}\left(\frac{q_1^2}{2} a \sqrt{ \mu T t}\right)}{a^2 q_1^4 + 4 q_2^2},
\end{equation}
which leads to the same asymptotics in the long-$t$ limit.

\subsection{Derivation of the response function}
\label{app:response_comb}
We start by Laplace transforming the corresponding Schwinger-Dyson equation~\eqref{eq:SD_response} with respect to both $t \rightarrow s$ and $t' \rightarrow s'$, finding
\begin{align}
    s R\qp(s,s') =& 
    -\mu q_\perp^2\,(T+u\q) R\qp(s,s')-\mu q_1^2\, \int \dperp{k} (T+u_{q_1,-\vb k_\perp})R_{(q_1,\vb k_\perp),\vb p}(s,s') \n\\
    &+ \frac{2\mu T}{s+s'} \left[ q_1^2\delta(q_1+p_1)+q_\perp^2\delta^d(\vb q+\vb p) \right].
            \label{appeq:response_00}
\end{align}
Note that the first Laplace transform, $t \rightarrow s$, acts on the first term in \cref{eq:SD_response} as 
\begin{equation}
    \partial_t R\qp(t,t') \qquad \rightarrow \qquad s R\qp(s,t') - R\qp(t=0^+,t')\Theta(-t'), 
\end{equation}
where the $\Theta(-t')$ stems from the causality of the response function --- see its definition in \cref{eq:def_response}. The second Laplace transform, $t' \rightarrow s'$, then suppresses the term proportional to $\Theta(-t')$.

Equation~\eqref{appeq:response_00} can be rewritten as
\begin{align}
    R\qp(s,s') = g_\perp(\vb q,s) \left\lbrace \frac{2\mu T}{s+s'} \left[ q_1^2\delta(q_1+p_1)+q_\perp^2\delta^d(\vb q+\vb p) \right]-\mu q_1^2 F\o(q_1;s,s') \right\rbrace,
    \label{appeq:response_0}
\end{align}
where we introduced the auxiliary function
\begin{equation}
    F\o(q_1;s,s') \equiv \int \dperp{q} (T+u_{q_1,-\vb q_\perp})R_{\vb q,\vb p}(s,s').
\end{equation}
Applying the operator $\int \dperp{q}(T+u_{q_1,-\vb q_\perp})$ to \cref{appeq:response_0} we thus obtain the self-consistency relation
\begin{align}
    F\o(q_1;s,s') =& \, M(q_1,s)\left[ \frac{2\mu T}{s+s'} q_\perp^2\delta^d(\vb q+\vb p) -\mu q_1^2 F\o(q_1;s,s') \right] \n\\
    &+ \frac{2\mu T}{s+s'} q_1^2\delta(q_1+p_1) (T+u\p) g_\perp(-\vb p,s).
    \label{appeq:F_self}
\end{align}
From now on, we will assume for simplicity $u(\vb x)$ to be rotationally invariant, so that $u\q$ depends only on $\abs{\vb q}=q_1^2+q_\perp^2$. This implies in particular
\begin{equation}
    g_\perp(\vb q,s)=g_\perp(-\vb q,s), \qquad M(q_1,s)= M(-q_1,s),
\end{equation}
as can easily be evinced from their definitions in \cref{eq:def_gperp,appeq:def_M}. Solving for $F\o$ in \cref{appeq:F_self} and replacing the result into \cref{appeq:response_0} then renders the response function given in \cref{eq:comb_response_laplace}, upon noting that
\begin{equation}
        1-\mu p_\perp^2(T+u\p) g_\perp(\vb p ,s) = s \, g_\perp(\vb p ,s),
        \label{appeq:relation}
\end{equation}
and by using the propagator $G\q(s)$ introduced in \cref{eq:comb_prop}.

Using the causality structure of the response function (see~c.f.~\ref{app:laplace_causality}), we can finally invert \cref{eq:comb_response_laplace} with $u(\vb x)=A\delta(\vb x)$ and $d=2$, which gives
\begin{align}
    R\qp(t,t')=& \, \frac{2\mu T \Theta(\tau) q_1^2\delta(q_1+p_1)}{(q_2^2-p_2^2)(4p_2^2+p_1^4)(4q_2^2+q_1^4)} \Bigg\lbrace  
    2q_2^3(4p_2^2+p_1^4) \mathrm e^{-q_2^2\mu\mathcal T \tau}\Big[2q_2\n\\
    &-q_1^2 \T{Erfi}\left(q_2 \sqrt{\mu\mathcal T \tau} \right) \Big]
    -(\vb p \leftrightarrow \vb q) 
    + \mathrm e^{\frac{q_1^4}{4}\mu\mathcal T \tau}(q_2^2-p_2^2)q_1^8 \T{Erfc}\left(\frac{q_1^2}{2}\sqrt{\mu\mathcal T \tau} \right) \Bigg\rbrace  
    \n\\&
    +2\mu T \Theta(\tau) p_2^2\delta(\vb q+\vb p) \mathrm e^{-q_2^2\mu\mathcal T \tau},
\end{align}
with $\tau=t-t'>0$ and $\cor T \equiv T+A$. 

\subsection{Derivation of the correlator for Dirichlet initial conditions}
\label{app:correlator_comb}

As in the previous sections, we start by Laplace-transforming the corresponding Schwinger-Dyson equation~\eqref{eq:SD_correlator} with respect to $t \rightarrow s$, finding
\begin{align}
        C\qp(s,t') =&\, g_\perp(\vb q,s) \Big[ C\qp(t=0^+,t')-\mu q_1^2\, \int \dperp{k} (T+u_{q_1,-\vb k_\perp})C_{(q_1,\vb k_\perp),\vb p}(s,t')\n\\
        &+ R\pq(t',s)\Big].
        \label{appeq:correlator_0}
\end{align}
Note that this time the quantity $C\qp(t=0^+,t')\neq 0$. On the contrary, this term can be used to enforce any particular initial condition at $t_0=0$, because by construction
\begin{equation}
    C\qp(t=t_0,t') = \phi\q(t_0)\phi\p(t_0)G\p(t'-t_0),
    \label{appeq:initial}
\end{equation}
where $G\q(t)$ is the propagator in \cref{eq:comb_prop}. We remark that the seemingly arbitrary choice $t_0=0$ is necessary to allow the use of the Laplace transform, but in fact it entails no loss of generality. Indeed, once a particular solution $C\qp(t,t')$ is found, one can simply introduce the new coordinates
\begin{equation}
    \tau=t+t_0, \qquad \tau'=t'+t_0,
\end{equation}
which practically amounts to shifting the initial conditions to the time $\tau=t_0$.
In particular, this may allow to find the correlator in the stationary state by eventually taking the limit $t_0\to -\infty$.

Upon further Laplace-transforming \cref{appeq:correlator_0} with respect to $t'\to s'$, we thus obtain
\begin{align}
    C\qp(s,s') =&\, g_\perp(\vb q,s) \Big[\phi\q\z\phi\p\z G\p(s') -\mu q_1^2\, \int \dperp{k} (T+u_{q_1,-\vb k_\perp})C_{(q_1,\vb k_\perp),\vb p}(s,s') \n\\
    &+R\pq(s',s) \Big],
    \label{appeq:correlator_1}
\end{align}
where we called for brevity $\phi\q\z \equiv \phi\q(t=0)$, and
where the response function $R\qp(s,s')$ was given in \cref{eq:comb_response_laplace} (note, however, the exchange of its arguments). 
Again, our strategy consists in integrating \cref{appeq:correlator_1} with respect to $\vb q_\perp$ to obtain a self-consistency equation. To this end, we first collect the various term in \cref{appeq:correlator_1} according to their dependence on $\vb q_\perp$, and use the property
\begin{equation}
    g_\perp(\vb q ,s)g_\perp(\vb q ,s') = \frac{g_\perp(\vb q ,s)-g_\perp(\vb q ,s')}{s'-s},
\end{equation}
which is readily proven by inspection of the definition~\eqref{eq:def_gperp} of $g_\perp(\vb q,s)$. We can thus rewrite
\begin{align}
    C\qp(s,s') =&\, g_\perp(\vb q,s)\phi\q\z\phi\p\z G\p(s') +  g_\perp(\vb q,s)g_\perp(\vb q,s') \delta^d(\vb q+\vb p) \frac{2\mu  T}{s+s'} p_\perp^2 \n\\
    &+ g_\perp(\vb q,s) \left[  \frac{2\mu  T}{s+s'} \frac{p_1^2\delta(q_1+p_1)\, s' \,G\p(s')}{s'-s} -\mu q_1^2\,F\t (q_1;s,s') \right]\n\\
    &-g_\perp(\vb q,s') \frac{2\mu  T}{s+s'} \frac{p_1^2\delta(q_1+p_1)\, s' \,G\p(s')}{s'-s} ,\label{appeq:Cqp_intermediate}
\end{align}
where we introduced the auxiliary function
\begin{equation}
    F\t (q_1;s,s')\equiv \int \dperp{q} (T+u_{q_1,-\vb q_\perp})C_{\vb q,\vb p}(s,s') .
\end{equation}
Applying $\int \dperp{q}(T+u_{q_1,-\vb q_\perp})$ to \cref{appeq:correlator_1} renders
\begin{align}
    F\t (q_1;s,s') =&\, \phi\p\z G\p(s')M\z(q_1,s)+  M(q_1,s)\left[ \mathcal A -\mu q_1^2\, F\t (q_1;s,s') \right] \n\\
    &-\cor A  M(q_1,s) + g_\perp(\vb p,s)g_\perp(\vb p,s') \delta (q_1+ p_1) \frac{2\mu  T}{s+s'} p_\perp^2\, (T+u\p),
    \label{appeq:self_F2}
\end{align}
where the functions $M\z,M$ were introduced in \cref{appeq:def_M,eq:def_M0}, respectively, and where we called for brevity
\begin{equation}
    \cor A \equiv \frac{2\mu  T}{s+s'} \frac{p_1^2\delta(q_1+p_1)\, s' \,G\p(s')}{s'-s}.
\end{equation}
Solving for $F\t$ in \cref{appeq:self_F2} gives
\begin{align}
    \frac{F\t (q_1;s,s')}{g_1(q_1,s)} =& \, 
    \phi\p\z G\p(s')M\z(q_1,s) + \mathcal A \left[ M(q_1,s)- M(q_1,s')  \right] \n\\
    &+  g_\perp(\vb p,s)g_\perp(\vb p,s') \delta (q_1+ p_1) \frac{2\mu  T}{s+s'} p_\perp^2\, (T+u\p),
    \label{appeq:solution_F2}
\end{align}
which can be further simplified by rewriting
\begin{align}
    &\mu p_\perp^2\, (T+u\p) = g^{-1}_\perp(\vb p,s)-s, \\
    &\mu q_1^2\, \left[ M(q_1,s)- M(q_1,s')  \right] = g^{-1}_1( q_1,s) - g^{-1}_1( q_1,s'),
\end{align}
as one can easily check starting from the definitions of $g_\perp,\, g_1$,~and~$M$ given in \cref{eq:def_gperp,eq:def_g1,appeq:def_M}, respectively. Replacing the expression of $F\t$ found in \cref{appeq:solution_F2} back into \cref{appeq:Cqp_intermediate} renders, after some algebra, the correlation function $C\qp(s,s')$ reported in \cref{eq:correlator_dirichlet}.

Note that specializing the latter to flat initial conditions $\phi\q\z\mapsto\phi\z$
yields
\begin{align}
    &\phi\p\z G\p(s')g_\perp(\vb q,s) \left[ \phi\q\z -\mu q_1^2 g_1(q_1,s)M\z(q_1,s) \right] \n\\
    &\mapsto  \left[ \phi\z\right]^2 G\p(s')g_\perp(\vb q,s) \left[ 1 -\mu q_1^2 g_1(q_1,s)M(q_1,s) \right] = \left[ \phi\z\right]^2 G\q(s)G\p(s'),
\end{align}
where the second step follows from the definitions of $M\z,M$ given in \cref{appeq:def_M,eq:def_M0}, while in the third step we recognized
\begin{equation}
    1-\mu q_1^2\,  g_1(q_1,s)  M (q_1,s) =   g_1(q_1,s),
\end{equation}
and we used the definition~\eqref{eq:comb_prop} of $G\q(s)$.

\subsection{Derivation of the equilibrium correlator}
\label{app:stationary_correlator_comb}

First, we note that simpler field theories generally allow to find the stationary correlator starting from the Dirichlet correlator computed in correspondence of \textit{fixed} initial conditions at time $t=t_0$. Indeed, if a time-domain expression of $C\qp(t,t')$ is available, one can formally send $t_0\to -\infty$, as we remarked at the beginning of \ref{app:correlator_comb}. We exemplify this route in \ref{app:toy_model} for the simple case of the free (Gaussian) field~\cite{Tauber}. However, directly inverting the Laplace expression of $C\qp(s,s')$ found in \cref{eq:correlator_dirichlet} proved challenging in the present context.

The strategy we adopt in this Section hinges instead on the assumption that the system reaches at long times the equilibrium distribution given in \cref{eq:equilibrium_dist_comb}. Here the fluctuation-dissipation theorem (in the form of \cref{eq:fdt}) links the stationary correlator $\cor C\qp(\tau)$ to the linear susceptibility
\begin{equation}
    \chi\qp(t,t') \equiv \eval{\fdv{\expval{\phi\q(t)}}{j_{-\vb p}(t')}}_{j=0},
    \label{appeq:susc_fourier}
\end{equation}
computed in the presence of a field $j(\vb x,t)$
linearly coupled to the Hamiltonian as in \cref{eq:field_hamiltonian_j}. The corresponding equation of motion follows from \cref{eq:equilibrium_dynamics} as
\begin{equation}
    \partial_t \phi(\vb x,t) = \div \left\lbrace \hat \mu(\vb x)\grad \left[  T  \phi(\vb x,t) +\int \dd{\vb y} u(\vb x-\vb y)\phi(\vb y) -j(\vb x,t)\right]+\bm \xi (\vb x,t) \right\rbrace.
\end{equation}
Taking its average, its spatial Fourier transform, and finally its functional derivative as in \cref{appeq:susc_fourier}, leads to the following evolution equation for the susceptibility:
\begin{align}
    \partial_t \chi\qp(t,t') =& 
    -\mu q_\perp^2\,(T+u\q) \chi\qp(t,t')-\mu q_1^2\, \int \dperp{k} (T+u_{q_1,-\vb k_\perp})\chi_{(q_1,\vb k_\perp),\vb p}(t,t')  \n \\
    &+ \mu  \delta(t-t') \left[ q_1^2\delta(q_1+p_1)+q_\perp^2\delta^d(\vb q+\vb p) \right]. 
    \label{appeq:SD_response}
\end{align}
The latter is formally analogous (up to a multiplicative factor $2T)$ to the evolution equation~\eqref{eq:SD_response} satisfied by the response function $R\qp(t,t')$, whose solution has been computed in \ref{app:response_comb}, hence we can write immediately its solution
\begin{equation}
    \chi \qp(s) = \frac{\mu }{s+s'}g_\perp(p_\perp,s) \left[ q_\perp^2\delta^d(\vb q+\vb p)+q_1^2\delta(q_1+p_1)\, s \,G\q(s) \right] \equiv \frac{\widetilde \chi \qp(s)}{s+s'}.
\end{equation}

Finally, by integrating the fluctuation-dissipation relation~\eqref{eq:fdt} one can obtain the stationary correlator $\cor C(s)$, provided that one first can fix the initial condition
\begin{equation}
    \cor C\qp(\tau=0^+) = \expval{\phi\q(t)\phi\p(t)}_{t\to \infty}.
\end{equation}
The latter encodes
the equal-time fluctuations of the field in the equilibrium state described by the distribution $\cor P[\phi]$ in \cref{eq:equilibrium_dist_comb}. Since $\cor P[\phi]$ features the Hamiltonian $\cor H[\phi]$ in \cref{eq:field_hamiltonian_j}, which is Gaussian, the equilibrium fluctuations of $\phi$ can be simply obtained by first constructing the generating functional \cite{LeBellac}
\begin{equation}
    \cor Z[j] = \int \cor D \phi \, \exp{ -\frac{1}{T} \cor H_j[\phi] } = \exp(\frac{1}{2T} \int \dslash{q}  \frac{j_{\vb q} j_{-\vb q}}{T+u\q} ),
    \label{appeq:gen_func}
\end{equation}
where
$\cor H_j[\phi]$ was introduced in \cref{eq:field_hamiltonian_j}, and
the functional integral in \cref{appeq:gen_func} is assumed to be normalized so that $\cor Z[j=0]=1$. We can then compute
\begin{equation}
    \expval{\phi\q \phi\p} =  \eval{T^2\frac{\delta^2 \cor Z[j]}{\delta j_{-\vb q} \delta j_{-\vb p}}}_{j=0}  =  \frac{T}{T+u\q} \delta^d(\vb q+\vb p),
\end{equation}
which is the result reported in \cref{eq:corr_initial}. Using the fluctuation-dissipation theorem written in the Laplace domain as in \cref{eq:fdt_laplace}, we then find the stationary correlator given in \cref{eq:equilibrium_correlator_comb}.

The latter can eventually be inverted to the time domain for selected choices of $u\q$. Note that the so-obtained $\cor C\qp(\tau)$ is only expected to reproduce the branch with $\tau>0$, while we physically expect it to be symmetric for $\tau<0$,~i.e.~$\cor C\qp(\tau)=\cor C\qp(\abs{\tau})$. The result depends in general on the spatial dimensionality, hence we find for instance in $d=2$, with $u(\vb x) = A\delta (\vb x)$,
\begin{align}
    \cor C\qp(\tau)
    =&\,  \frac{T}{\cor T}\delta(\vb q+\vb p) \mathrm e^{-q_2^2\mu\mathcal T \tau}  - 
    \frac{T}{\cor T} 
    q_1^2\delta(q_1+p_1) \Bigg\lbrace  
    \frac{2q_2 \mathrm e^{-q_2^2\mu\mathcal T \tau}\left[2q_2-q_1^2 \T{Erfi}\left(q_2 \sqrt{\mu\mathcal T \tau} \right) \right]}{(p_2^2-q_2^2)(4q_2^2+q_1^4)}
     \n\\
    & +(\vb q \leftrightarrow \vb p)   + \frac{4q_1^4 \mathrm e^{\frac{q_1^4}{4}\mu\mathcal T \tau} \T{Erfc}\left(\frac{q_1^2}{2}\sqrt{\mu\mathcal T \tau} \right)}{(4p_2^2+p_1^4)(4q_2^2+q_1^4)} \Bigg\rbrace  ,
\end{align}
where we called again $\cor T \equiv T+A$.

As we stressed, the derivation proposed in this Section relied on the assumption that the system reaches thermal equilibrium described by the Boltzmann distribution. In the next Section we thus prove that the equilibrium solution represents the only admissible time-translational invariant solution for this system.

\subsection{Proof of the uniqueness of the stationary solution}
\label{app:uniqueness}
In \cref{sec:equilibrium}, the sole requirement of time-translational invariance led to the condition~\eqref{appeq:double_SD} for the correlator $C\qp(t,t')$.  
For simpler field theories, this condition is sufficient to construct explicitly the stationary correlator --- we exemplify this for the free field in \ref{app:toy_model}. However, in the case of the comb the integral equation~\eqref{appeq:double_SD} is hard to solve as it stands, and thus we proceed differently.

First, we use the causality structure of the response function and of the stationary correlator to reduce the time dependence in \cref{appeq:double_SD} to a single variable. To this end, we step to the Laplace domain, where the response functions in \cref{appeq:source} take the form
\begin{equation}
    \cor S\qp(s,s') = \frac{\widetilde R\qp(s) +\widetilde R\pq(s')}{s+s'} = \frac{\widetilde R\qp(s) +\widetilde R\qp(s')}{s+s'}.
\end{equation}
In the first step we used the causal form of the response function $R\qp(s,s')$ reported in \cref{eq:response_causal}, while in the second step we used the symmetry under $\vb q \leftrightarrow \vb p$ of the particular solution found in \cref{eq:comb_response_laplace}. Similarly, the correlator $C\qp(s,s')$ can be written in terms of $\cor C\qp(s)$ as in \cref{eq:corr_laplace_symmetry}. By plugging this decomposition into \cref{appeq:double_SD} we obtain two distinct (and equivalent) relations, depending either on $s$ or $s'$, the first of which reads
\begin{equation}
    \cor C\qp(s) = \cor G\qp \left\lbrace \widetilde R \qp(s) -\mu  \int \dperp{k} (T+u_{q_1,-\vb k_\perp})\left[q_1^2\,\cor C_{(q_1,\vb k_\perp),\vb p}(s) +p_1^2\,\cor C_{\vb q,(p_1,\vb k_\perp)}(s) \right]\right\rbrace .
    \label{appeq:double_SD_2}
\end{equation}
Next, we note that \cref{appeq:double_SD_2} fixes completely the \textit{diagonal} part of the correlator $\cor C\qp(s)$ in momentum space. Indeed, let us decompose (without loss of generality)
\begin{equation}
    \cor C\qp(s) = c\q\o(s) \delta^d(\vb q+\vb p) + c\t\qp(s) \delta(q_1+p_1).
    \label{appeq:decomposition_1}
\end{equation}
We have additionally assumed that $\cor C\qp(s) \propto \delta(q_1+p_1)$, as expected from the translational invariance of the system along the backbone.
Plugging this form into \cref{appeq:double_SD_2} renders
\begin{align}
    c\q\o(s) =&\, \frac{T}{T+u\q}g_\perp(\vb q,s), \label{appeq:c1} \\
    c\t\qp(s) =& \, \mu q_1^2 \, \cor G\qp \Bigg\lbrace 2 T g_\perp(\vb p,s)  s G\q(s) -  \left[ (T+u\q)c\q\o(s)+(T+u\p)c\p\o(s) \right] \n \\
    & - \int \dslash{k} \delta(k_1-q_1) (T+u_{-\vb k})\left[ c\t_{\vb k,\vb p}(s) +  c\t_{\vb q,\vb k}(s) \right] \Bigg\rbrace . \label{appeq:c2}
\end{align}
Note that the diagonal dependence encoded in $c\q\o(s)$ is exactly the same as that of the equilibrium correlator reported in \cref{eq:equilibrium_correlator_comb}. This suggests to express the off-diagonal part $c\t\qp(s)$ as a deviation with respect to the equilibrium case: we thus introduce
\begin{equation}
    c\t\qp(s) = -\mu q_1^2 T g_\perp(\vb p,s)G\q(s)+ c\tr\qp(s).
\end{equation}
We then plug this into \cref{appeq:c2}, and use \cref{appeq:c1} to recognize
\begin{align}
    \left[ (T+u\q)c\q\o(s)+(T+u\p)c\p\o(s) \right] &= T \left[ g_\perp(\vb q,s)+g_\perp(\vb p,s) \right]\n\\
    &= g_\perp(\vb q,s)\,g_\perp(\vb p,s)\,\left[ 2s+\cor G\qp^{-1}\right],
\end{align}
where in the last step we used the definition of $g_\perp$ given in \cref{eq:def_gperp}, and that of $\cor G\qp$ in \cref{appeq:def_Gqp}. This leads, after some algebra, to a simpler relation of the form
\begin{equation}
    c\tr\qp(s) =- \mu q_1^2 \, \cor G\qp  \int \dslash{k} \delta(k_1-q_1) (T+u_{-\vb k})\left[ c\tr_{\vb k,\vb p}(s) +  c\tr_{\vb q,\vb k}(s) \right]  . 
    \label{appeq:c3}
\end{equation}
The latter integral equation is homogeneous, and thus it is clearly
solved by $c\tr\qp(s) \equiv 0$. In the following, we will prove that such solution is unique: as a consequence,
the equilibrium correlator given in \cref{eq:equilibrium_correlator_comb} is the unique solution of \cref{appeq:double_SD_2}, which defines the \textit{stationary} correlator. To this end, we introduce another transformation as
\begin{equation}
    c\f\qp(s) \equiv \cor G\qp^{-1} \, c\tr\qp(s),
\end{equation}
in terms of which \cref{appeq:c3} becomes
\begin{equation}
    c\f\qp(s) =- \mu q_1^2 \,   \int \dslash{k} \delta(k_1-q_1) (T+u_{-\vb k})\left[ \cor G_{\vb k,\vb p} c\f_{\vb k,\vb p}(s) + \cor G_{\vb q,\vb k} c\f_{\vb q,\vb k}(s) \right]  . 
    \label{appeq:c4}
\end{equation}
This form makes clear that the dependence of $c\f\qp(s)$ on $\vb q$ and $\vb p$ is \textit{additive}, namely
\begin{equation}
    c\f\qp(s) = f\q(s) + f\p(s),
\end{equation}
for some unknown function $f\q(s)$. Plugging this decomposition into \cref{appeq:c4} and introducing the auxiliary function
\begin{equation}
    N\q \equiv \mu q_1^2 \,   \int \dslash{k} \delta(k_1-q_1) (T+u_{-\vb k})\cor G_{\vb k,\vb q},
\end{equation}
we obtain (using the parity of $\cor G\qp$)
\begin{align}
    &\left[\left( 1+N\q\right) f\q(s) + \mu q_1^2 \,   \int \dslash{k} \delta(k_1-q_1) (T+u_{-\vb k})\cor G_{\vb k,\vb q} f\k(s) \right]\n \\
    &= - \left[ \left( 1+N\p\right) f\p(s) + \mu q_1^2 \,   \int \dslash{k} \delta(k_1-q_1) (T+u_{-\vb k})\cor G_{\vb k,\vb p} f\k(s) \right].
    \label{appeq:separation}
 \end{align}
The left-hand side of this equation depends on $\vb q_\perp$, whereas the right-hand side depends on $\vb p_\perp$ --- hence they must be equal to a quantity that is independent of both $\vb q_\perp$ and $\vb p_\perp$:
\begin{align}
    &\left( 1+N\q\right) f\q(s) + \mu q_1^2 \,   \int \dslash{k} \delta(k_1-q_1) (T+u_{-\vb k})\cor G_{\vb k,\vb q} f\k(s) = a(q_1,s),\\
    &\left( 1+N\p\right) f\p(s) + \mu q_1^2 \,   \int \dslash{k} \delta(k_1-q_1) (T+u_{-\vb k})\cor G_{\vb k,\vb p} f\k(s) = -a(q_1,s).
\end{align}
By symmetry under the exchange $\vb q_\perp \leftrightarrow \vb p_\perp$, we argue that it can only be $a(q_1,s)=0$.
This way we have reduced \cref{appeq:double_SD} to the single-variable integral equation
\begin{equation}
    \left( 1+N\q\right) f\q(s) + \mu q_1^2 \,   \int \dslash{k} \delta(k_1-q_1) (T+u_{-\vb k})\cor G_{\vb k,\vb q} f\k(s) = 0.
\end{equation}
Taking another convolution against $(T+u_{-\vb q})$ finally renders
\begin{equation}
    \int \dslash{q} (T+u_{-\vb q}) (1+2 N\q)f\q(s)  = 0,
\end{equation}
which is only satisfied if $f\q(s)=0$. This concludes the proof of the uniqueness.

\subsection{Double Laplace transform of causal/stationary functions}
\label{app:laplace_causality}
The causality structure of a function in the time domain carries signatures on its Laplace transform. Indeed, it is straightforward to prove that~\cite{Debnath2015,ditkin2017operational,Wald2021}
\begin{equation}
    F(t_1,t_2)=f(t_2-t_1)\Theta(t_2-t_1) \qquad \rightarrow \qquad \hat F(s_1,s_2) = \frac{\hat f(s_2)}{s_1+s_2}, \label{appeq:laplace_causal}
\end{equation}
and similarly
\begin{align}
    &G(t_1,t_2)=g(\abs{t_2-t_1})=g(t_2-t_1)\Theta(t_2-t_1)+g(t_1-t_2)\Theta(t_1-t_2)\n\\
    &\rightarrow \qquad \hat G(s_1,s_2) = \frac{\hat g(s_1)+\hat g(s_2)}{s_1+s_2}.
\end{align}
This simplifies significantly the inverse Laplace transformation of the response function $R\qp(s,s')$ and the stationary correlator $\cor C\qp(s,s')$ derived in \ref{app:response_comb} and \ref{app:stationary_correlator_comb}, respectively:
we basically have to transform only with respect to $s$, and not with respect to both $s$ and $s'$. Unfortunately this is not the case for the correlator $C\qp(s,s')$ in \cref{eq:correlator_dirichlet} for Dirichlet initial conditions, because the corresponding $C\qp(t,t')$ is not expected to be time-translational invariant.

\subsection{Particles in homogeneous space: analysis of a fluctuating Gaussian field in the Laplace domain}
\label{app:toy_model}

\rev{Here we consider the relaxational dynamics of a Gaussian field~\cite{Tauber},
\begin{gather}
    \dot{\phi}_{\vb q}(t) = -\alpha\q \phi_{\vb{q}}(t) + \nu_{\vb q}(t)  ,
    \label{chap1:eq:fieldFourier} \\
    \expval*{\nu_{\vb q}(t)\nu_{\vb q'}(t')}= \Omega\q \delta^d(\vb q+\vb q')\delta(t-t') ,
    \label{chap1:eq:field_noise_Fourier}
\end{gather}
to which \cref{eq:field_comb} for the fluctuating particle density essentially reduces 
in homogeneous space (i.e.~for $\hat \mu(\vb x)\equiv \mu \,\mathbb{1}_{d}$), upon identifying $\alpha\q= \mu q^2(T+u\q)$ and $\Omega\q=2\mu Tq^2$.}
Although this problem is easily solvable in the time domain, here we will characterize it instead in the Laplace domain. This serves as a benchmark for the methods employed above in the case of the comb, which in contrast cannot be analyzed directly in the time domain.

First, the propagator corresponding to \cref{chap1:eq:fieldFourier} can be immediately checked to give
\begin{equation}
    G\q(s) = \frac{1}{s+\alpha\q} \qquad \leftrightarrow \qquad G\q(t) = \Theta(t) \mathrm e^{-\alpha\q t}.
\end{equation}
To find the correlation function, we first write the Schwinger-Dyson equations (see \cref{sec:twopoint-general})
\begin{align}
    \partial_t C\qp(t,t') =& 
    -\alpha\q C\qp(t,t')    + R\pq(t',t) , \label{appeq:FF_SD_corr}\\
    \partial_t R\qp(t,t') =& 
    -\alpha\q R\qp(t,t')
    + \Omega\q \delta^d(\vb q+\vb p)\delta(t-t'). 
\end{align}
Solving the latter in the Laplace domain (as we did in \ref{app:response_comb} for the comb) yields
\begin{equation}
    R\qp(s,s') = \frac{\Omega\q \delta^d(\vb q+\vb p)}{(s+s')(s+\alpha\q)} = \frac{\Omega\q \delta^d(\vb q+\vb p)}{(s+s')}G\q(s),
    \label{appeq:FF_response}
\end{equation}
which can be inverted to the time domain (with respect to both its Laplace variables) to give
\begin{equation}
    R\qp(t,t') = \Omega\q \delta^d(\vb q+\vb p) \Theta(t') \Theta(t-t') \mathrm e^{-\alpha\q (t-t')}.
\end{equation}
To find the correlator, we Laplace-transform \cref{appeq:FF_SD_corr} and plug in the solution~\eqref{appeq:FF_response} for the response function:
\begin{equation}
    C\qp(s,s') = \frac{1}{s+\alpha\q} \left[ C\qp(t=0^+,s')+  \frac{\Omega\q \delta^d(\vb q+\vb p)}{(s+s')(s'+\alpha\q)} \right].
\end{equation}
Transforming back to the time domain yields
\begin{equation}
    C\qp(t,t') = C\qp(t=0^+,t') \mathrm e^{-\alpha\q t} + \frac{\Omega\q \delta^d(\vb q+\vb p) }{2\alpha\q} \left[ \mathrm e^{-\alpha\q |t-t'|}-\mathrm e^{-\alpha\q (t+t')}\right],
    \label{appeq:FF_corr_0}
\end{equation}
where the term $C\qp(t=0^+,t')$ must be fixed self-consistently by using the initial conditions
\begin{equation}
    C\qp(t=t_0,t') = \phi\q(t_0) \phi\p(t_0) G\p(t'-t_0),
\end{equation}
specified in correspondence of $t_0<t'$. Indeed, computing \cref{appeq:FF_corr_0} in $t=t_0$ and solving for $C\qp(t=0^+,t')$ gives
\begin{equation}
    C\qp(t=0^+,t') =\phi\q(t_0) \phi\p(t_0) \mathrm e^{-\alpha\q (t'-2t_0)} + \frac{\Omega\q \delta^d(\vb q+\vb p) }{2\alpha\q}  \mathrm e^{-\alpha\q t'} \left( 1 -\mathrm e^{2\alpha\q t_0}\right),
\end{equation}
which can then be replaced back into \cref{appeq:FF_corr_0} to find the Dirichlet correlator~\cite{Tauber}
\begin{equation}
    C\qp(t,t') = \phi\q(t_0) \phi\p(t_0) \mathrm e^{-\alpha\q (t+t'-2t_0)} + \frac{\Omega\q \delta^d(\vb q+\vb p) }{2\alpha\q} \left[ \mathrm e^{-\alpha\q |t-t'|}-\mathrm e^{-\alpha\q (t+t'-2t_0)}\right].
    \label{appeq:FF_dirichlet}
\end{equation}
Note that, starting from the flat initial condition $\phi\q(0)=0$, the Dirichlet correlator reads in the Laplace domain
\begin{equation}
    C\qp(s,s') =  \frac{\Omega\q \delta^d(\vb q+\vb p)}{(s+s')(s+\alpha\q)(s'+\alpha\q)} .
\end{equation}
Note also that, unfortunately, an explicit Laplace expression of the correlator starting from initial conditions set at a generic time $t_0$ is not available. Indeed, $t=0$ plays a special role for the Laplace transform, which in some sense breaks time-translational invariance. This is why in \ref{app:correlator_comb} we had to fix the initial conditions at time $t_0=0$.

Finally, to find the stationary correlator in the Laplace domain we start by writing another Schwinger-Dyson equation, akin to \cref{appeq:FF_SD_corr}:
\begin{equation}
    \partial_{t'} C\qp(t,t') = 
    -\alpha\p C\qp(t,t')    + R\qp(t,t').
\end{equation}
We then impose the stationarity condition
\begin{equation}
    (\partial_t+\partial_{t'})C\qp(t,t')  \equiv 0 \qquad \rightarrow \qquad C\qp(t,t') = \frac{R\qp(t,t')+R\pq(t',t)}{\alpha\q+\alpha\p}.
\end{equation}
Plugging in the Laplace expression~\eqref{appeq:FF_response} of the response function yields
\begin{equation}
    C\qp(s,s') = \frac{\Omega\q \delta^d(\vb q+\vb p)}{2\alpha\q (s+s')} \left( \frac{1}{s+\alpha\q}+\frac{1}{s'+\alpha\q}\right),
\end{equation}
whose time-domain version
\begin{equation}
    C\qp(t,t') = \frac{\Omega\q \delta^d(\vb q+\vb p) }{2\alpha\q}  \mathrm e^{-\alpha\q |t-t'|}
\end{equation}
clearly coincides with the formal limit $t_0\to -\infty$ taken in \cref{appeq:FF_dirichlet}. It also coincides with the \textit{equilibrium} correlator obtained by using the fluctuation-dissipation theorem~\cite{Tauber,Venturelli_2022_2parts}.

\section{Structure factor and intermediate scattering function}
\label{app:structure}
In this appendix we recall some basic concepts from the physics of liquids~\cite{McDonald_book}, so as to make contact with the notation used in SDFT.

Let us first introduce the (static) structure factor
\begin{equation}
    S(\mathbf{q}) = \frac{1}{N} \sum_{i,j=1}^N \langle \mathrm e^{-i \mathbf{q} \cdot (\mathbf{r}_i - \mathbf{r}_j)} \rangle = \frac{1}{N} 
    \expval{\rho\q \rho_{-\vb q}},
    \label{appeq:structure}
\end{equation}
where in the second step we used the definition of the density field $\rho$ given in \cref{eq:density}. First, note that one can always decompose \cref{appeq:structure} as
\begin{equation}
    S(\mathbf{q}) = 1+ \frac{1}{N} \sum_{i\neq j} \langle \mathrm e^{-i \mathbf{q} \cdot (\mathbf{r}_i - \mathbf{r}_j)} \rangle ,
    \label{appeq:structure2}
\end{equation}
where the first term accounts for self-correlations.
For large $\vb q$, the second term rapidly oscillates and averages out to zero, so that $S(\vb q)\sim 1$; moreover, from \cref{appeq:structure} it follows that $S(\vb q= \bm 0)=N$.
As such, by construction, $S(\vb q) $ does not admit an inverse Fourier transform.
Using \cref{appeq:structure} and the definition of the density fluctuation $\phi$ introduced in \cref{eq:fluctuation}, we can rewrite
\begin{align}
    S(\vb q) &= \rho_0 \left[\delta^d(\vb q) + \frac{1}{N}\expval{\phi\q \phi_{-\vb q}} \right] \n\\
    &= 1+\rho_0 \delta^d(\vb q) +\rho_0 h(\vb q).
    \label{appeq:h(q)}
\end{align}
Above, in the first line we used $[\delta^d(\vb q)]^2= \cor V \delta^d(\vb q)$, where $\cor V$ denotes tbe volume of the system (i.e.~$\rho_0=N/\cor V$),
while the second line defines the \textit{pair correlation function} $h(\vb q)$ ---
note that the latter \textit{does} admit an inverse Fourier transform~\cite{Demery2014,McDonald_book}. Note also that the term $\rho_0 \delta^d(\vb q)$ in \cref{appeq:h(q)} is only due to the presence of a uniform nonzero background density $\rho_0$; since this only entails a difference in $\vb q=\bm 0$, one typically rather plots 
\begin{equation}
    \mathfrak S(\vb q) = S(\vb q) - \rho_0 \delta^d(\vb q) = \frac{\rho_0}{N}\expval{\phi\q \phi_{-\vb q}}.
\end{equation}
Using \cref{eq:corr_initial}, we deduce that in our case
\begin{equation}
    \mathfrak S(\vb q) = \frac{\rho_0}{N} \cor C_{\vb q, -\vb q}(\tau=0^+) = \frac{T}{T+u\q},
\end{equation}
where again we used $\delta^d(\vb q=\bm 0)=\cor V$. As noted in earlier works~\cite{Demery2014,dean2018stochastic}, this result --- stemming from the linearization of the Dean--Kawasaki equation that we performed in \cref{sec:dean} --- corresponds to the random phase (or Debye-H\"uckel) approximation in the physics of liquids or electrolytes~\cite{McDonald_book}. Under such approximation, we note that the property $\mathfrak S(\bm 0)=0$ is lost, whereas $\mathfrak S(\vb q \to \infty)=1$, as expected.

Similarly, the temporal evolution of density fluctuations can be described using the intermediate scattering function~\cite{McDonald_book}
\begin{equation}
    F(\mathbf{q},t) = \frac{1}{N} \sum_{i,j=1}^N \langle \mathrm e^{-i \mathbf{q} \cdot (\mathbf{r}_i(t) - \mathbf{r}_j(0))} \rangle = \frac{1}{N} 
    \expval{\rho\q(t) \rho_{-\vb q}(0)}.
    \label{appeq:ISF}
\end{equation}
Similar considerations to those spelled out above for the structure factor apply also here, so that in practice it is useful to subtract the constant background and plot instead
\begin{equation}
    \mathfrak F(\vb q, t) = F(\vb q,t) - \rho_0 \delta^d(\vb q) = \frac{\rho_0}{N}\expval{\phi\q(t) \phi_{-\vb q}(0)} = \frac{\rho_0}{N} \cor C_{\vb q, -\vb q}(t),
    \label{eq:ISF-def}
\end{equation}
where in the last step we recognized the two-point function introduced in \cref{eq:comb_correlator_def,eq:property_correlator}. For the comb geometry, the latter is only known explicitly in the Laplace domain, i.e.~$\cor C\qp(s)$ given in \cref{eq:equilibrium_correlator_comb}. However, its form suggests to rescale the momenta as $z_1^2 = q_1^2/ \sqrt{s}$ and $z_\perp^2 = q_\perp^2/ s$; in particular, the Fourier transform of the interaction potential $u(\vb x)$ this way reads 
\begin{equation}
    u\q = u_{(z_1 s^{1/4},z_\perp s^{1/2})}.
    \label{eq:changeofvars}
\end{equation}
Upon formally taking the inverse Laplace transform, it then becomes evident that $C_{\vb q, -\vb q}(t)$ collapses, for large times $t$, on a scaling function that does not depend on the details of the interaction potential $u\q$, but merely on $u_{\vb q=\bm 0}$ --- see \cref{eq:changeofvars} with $s\sim 0$. As noted in \ref{app:stationary_correlator_comb}, this is precisely the limit in which the inverse Laplace transform can be computed explicitly: this way we find, for $d=2$,
\begin{align}
    \mathfrak F(\vb q, t) &\underset{t \gg 1}{\sim}  \mathfrak F(y_1= q_1 t^\frac14, y_2=q_2 t^\frac12) \label{eq:scaling-ISF} \\
    &= \frac{T}{\cor T} \mathrm e^{-\mu \cor T y_2^2 } 
    \Bigg\lbrace 1+
    \frac{4y_1^2}{L(y_1^4 + 4 y_2^2)} 
    \Bigg[\frac{    y_1^4}{ (y_1^4 + 4 y_2^2)}     \left(1-
         \mathrm e^{ \mu \cor T (  y_2^2+y_1^4/4)}  \text{Erfc} \Big( \frac{y_1^2}{2}  \sqrt{ \mu\cor T} \Big) \right)
    \n\\ &
     \quad -     y_2^2  \mu \cor T - \frac{ \mathrm e^{\mu \cor T y_2^2 } y_1^2  \sqrt{\mu \cor T}}{2\sqrt{\pi}  }+
     \frac{ y_1^2\left ( 4 y_2^2 -y_1^4 + 2  y_2^2 (y_1^4 + 4 y_2^2) \mu \cor T\right)}{ y_2 (y_1^4 + 4 y_2^2)}  \text{Erfi}(4y_2 \sqrt{\mu \cor T})  \Bigg] \Bigg\rbrace,
    \n
\end{align}
where we called $\cor T = T+u_{\vb q=\bm 0}$. This function is plotted in \cref{fig:ISF}.
In particular, note that the term $\propto \delta(q_1 + p_1)$ in \cref{eq:equilibrium_correlator_comb} gives rise to a term $\propto L$ when computed for $\vb p=\vb q$ (as prescribed by \cref{eq:ISF-def}), where $L$ is the linear size of the system, whereas the term $\propto \delta^d(\vb q+\vb p)$ gives rise to a term $\propto \cor V=L^d$. Intuitively, this is because the structure factor and the ISF defined above entail averages over all particles in the system, among which only a subextensive fraction resides on the backbone. As a consequence, the correction to the ISF due to the particles subdiffusing along the backbone appears as a subleading contribution in \cref{eq:scaling-ISF}, which is expected to vanish in the thermodynamic limit $\cor V\to\infty$ (while it remains relevant in a finite system, see \cref{fig:ISF}).

\section{Details on the generalized correlation profiles}
\label{app:constrained_profiles}
Here we detail the derivation of the generalized correlation profiles between the tracer position and the density of all other bath particles, discussed in \cref{sec:general_prof_constrained}. 

Several steps in the following derivation resemble the ones carried out in Sec.~II.D of the Supplemental Material of Ref.~\cite{venturelli2024universal}, to which we refer the reader. In particular, the profiles $\psi\q$ and $g\q$ introduced in \cref{eq:hq_def,eq:gq_def} can be computed directly by using these very same methods, starting from the coupled equations of motion~\eqref{eq:tracer_fourier_comb}~and~\eqref{eq:field_fourier_comb}
for the tracer and the bath density. 
In the following, we limit ourselves for simplicity to the derivation of the generating function $w\q(\lambda,t)$ introduced in \cref{eq:wq_def}, from which the previous two profiles can actually be generated.

\subsection{Derivation of the generating function}
\label{app:wq}
We start from the definition~\eqref{eq:wq_def}
of
the generating function $w\q(\lambda,t)$,
and note that
\begin{equation}
    w_{\vb q}(\lambda, t) = \frac{\expval{  \phi_{\vb q}(t) \mathrm e^{ ( \lambda +i q_1)   r_0(t)}}}{ \expval{\mathrm e^{  \lambda   r_0(t)}} }= \frac{\expval{  \phi_{\vb q}(t) \mathrm e^{ ( \lambda +i q_1)   r_0(t)}}}{ \expval{\mathrm e^{  \lambda   r_0(t)}}_0 } +\order{h^2},
    \label{appeq:w}
\end{equation}
meaning that the leading-order contribution to $w_{\vb q}( \lambda, t)$ can be captured by computing the denominator only up to $\order{h^0}$. Since \cref{eq:tracer_fourier_comb} reduces to $1d$ Brownian motion if $h=0$, such denominator simply reads~\cite{Tauber}
\begin{equation}
    \expval{\mathrm e^{\lambda \, r_0(t)}}_0 =  
    \mathrm e^{\lambda^2 \mu_0 T t}.
    \label{appeq:cgf_0}
\end{equation}
At long times, we expect $w_{\vb q}( \lambda,t)$ to reach a stationary state satisfying 
\begin{align}
    0\equiv  \partial_t w_{\vb q}( \lambda) 
    =  \frac{(\partial_t-\lambda^2 \mu_0 T
    )\expval{  \phi_{\vb q}(t) \mathrm e^{ ( \lambda +i q_1)   r_0(t)}}}{ \expval{\mathrm e^{  \lambda \, r_0(t)}}_0 } +\order{h^2} ,
    \label{appeq:wq_stat}
\end{align}
where we used \cref{appeq:cgf_0,appeq:w}. 
(Note that the latter is an \textit{assumption}, to be verified a posteriori.)
We then focus on
\begin{equation}
    \partial_t \expval{  \phi_{\vb q}(t) \mathrm e^{ ( \lambda +i q_1)   r_0(t)}} = \expval{ \dot \phi_{\vb q}(t) \mathrm e^{ ( \lambda +i q_1)   r_0(t)}} +(\lambda+iq_1) \expval{ \dot r_0(t)  \phi_{\vb q}(t) \mathrm e^{ ( \lambda +i q_1)   r_0(t)}},
    \label{appeq:pieces}
\end{equation}
which we obtained by using Stratonovich calculus.
Indeed, note that
both noise terms in \cref{eq:tracer_fourier_comb,eq:field_fourier_comb}
are additive, hence the result of the calculation does not depend on the choice of the stochastic calculus convention.
The two terms on the r.h.s.~of \cref{appeq:pieces} can then be computed by using the coupled equations of motion;
in particular, the first term gives
\begin{align}
    \frac{\expval{  \dot \phi_{\vb q}(t) \mathrm e^{ ( \lambda +i q_1)   r_0(t)}}}{ \expval{\mathrm e^{  \lambda   r_0(t)}}_0 } =& -\mu q_\perp^2 (T+u\q) w\q(\lambda)  -\mu q_1^2 \int \dperp{q} (T+u\q) w\q(\lambda)  \n\\
    &-h\mu \left[ q_\perp^2 v\q +q_1^2 v_{q_1}(\vb x_\perp=\bm 0)\right],
    \label{appeq:block1}
\end{align}
because by using the Novikov theorem \cite{Novikov_1965,Luczka_2005} one can prove that~\cite{venturelli2024universal}
\begin{equation}
    \expval{\nu\q(t) \mathrm e^{(\lambda+iq_1)r_0(t)}}=0.
\end{equation}
Similarly, the second term yields
\begin{align}
    &\frac{\expval{  \dot r_0(t) \phi_{\vb q}(t) \mathrm e^{ ( \lambda +i q_1)   r_0(t)}}}{ \expval{\mathrm e^{  \lambda   r_0(t)}}_0 } 
    \n\\ &=
    \frac{ \expval{\eta_0(t) \phi\q(t) \mathrm e^{ ( \lambda +i q_1)   r_0(t)}  } }{ \expval{\mathrm e^{  \lambda   r_0(t)}}_0 }
    -h\mu_0 \int \dslash{p} ip_1 v\p \frac{ \expval{\phi\p(t) \phi\q(t) \mathrm e^{ ( \lambda +i q_1+ip_1)   r_0(t)}  } }{ \expval{\mathrm e^{  \lambda   r_0(t)}}_0 } 
    \n\\
    &= 
    (iq_1+\lambda ) \mu_0 T w\q(\lambda) +iq_1 h\mu_0 T \frac{v_{-\vb q}}{T+u\q}.
    \label{appeq:block2}
\end{align}
Above we noted that
\begin{align}
    \expval{\phi\p(t) \phi\q(t) \mathrm e^{ ( \lambda +i q_1+ip_1)   r_0(t)}  } &= \expval{\phi\p(t) \phi\q(t)}_0 \expval{ \mathrm e^{ ( \lambda +i q_1+ip_1)   r_0(t)}  }_0 +\order{h} \n\\
    &= \frac{T}{T+u\q}\delta^d(\vb q+\vb p) \expval{\mathrm e^{  \lambda   r_0(t)}}_0 +\order{h},
    \label{appeq:phi_phi_exp}
\end{align}
where
in the first step we used that the stochastic processes $r_0(t)$ and $\phi\q(t)$ are independent at $\order{h^0}$ and thus they factorize, while in the second step we inserted the stationary two-point function of the field found in \cref{eq:corr_initial}. 
Moreover, the expectation value involving the noise $\eta_0(t)$ can be derived as
in~\cite{venturelli2024universal} by using the Novikov theorem, yielding
\begin{equation}
    \expval{\eta_0(t) \phi\q(t) \mathrm e^{ ( \lambda +i q_1)   r_0(t)}  } = (iq_1+\lambda) \mu_0 T  \expval{ \phi\q(t) \mathrm e^{ ( \lambda +i q_1)   r_0(t)}  }.
\end{equation}
Plugging \cref{appeq:block1,appeq:block2} back into \cref{appeq:wq_stat} yields the defining equation of the stationary $w\q(\lambda)$, namely
\begin{align}
    G_0^{-1}(\vb q,\lambda) w\q(\lambda) =&  -\mu q_1^2 \int \dperp{q} (T+u\q) w\q(\lambda)  -h\mu \left[ q_\perp^2 v\q 
    +q_1^2 v_{q_1}(\vb x_\perp=\bm 0)\right] \n\\&+iq_1 h\mu_0 T \frac{v_{-\vb q}}{T+u\q},
    \label{appeq:wq_implicit}
\end{align}
where the function $ G_0(\vb q,\lambda)$ was introduced in \cref{eq:G0}.
Equation \eqref{appeq:wq_implicit} can as usual be solved self-consistently by applying $\int \dperp{q}(T+u\q)$ to both its sides; a tedious but straightforward calculation then renders the expression of $w\q(\lambda)$ reported in \cref{eq:wq}.

\subsection{Derivation and large-distance behavior of correlation profiles}
Other generalized correlation profiles can be promptly generated starting from $w\q(\lambda)$ given in \cref{eq:wq}. For instance, the profile $\psi\q$ introduced in \cref{eq:hq_def} can be simply computed as $\psi\q = w\q(\lambda=0)$. Similarly, since $g\q = \eval{\dv{w\q(\lambda)}{\lambda}}_{\lambda=0}$ (see \cref{eq:gq_def}), by deriving implicitly \cref{appeq:wq_implicit} we obtain
\begin{equation}
    G_0^{-1}(\vb q,\lambda=0) \,g\q =  -\mu q_1^2 \int \dperp{q} (T+u\q) g\q  +iq_1 \mu_0 T \left[\frac{hv_{-\vb q}}{T+u\q} +2\psi\q  \right].
    \label{appeq:gq_implicit}
\end{equation}
The self-consistent solution of \cref{appeq:gq_implicit} renders the stationary profile $g\q$, which is reported in \cref{eq:gq_nof}. 
Its large-distance behavior can then be inspected e.g.~by using the method presented in Sec.~I.C of the Supplementary Material in Ref.~\cite{venturelli2024universal}, or more simply by formally replacing the interaction potential by $u(\vb x)\mapsto u_{\vb q=\bm 0} \delta(\vb x)$ (indeed, note that we have assumed everywhere the interaction potential $u(\vb x)$ to decay rapidly, so that it admits a Fourier transform). Either way leads to the asymptotic result reported in \cref{eq:g-large-distance}.

\subsection{On higher-order cumulants}
\label{app:uniform_kurtosis}
In \cref{sec:MGF} we noted that our prediction for the cumulant generating function~\eqref{eq:logPsi} of the tracer position seems to imply that all cumulants of $r_0(t)$ grow linearly with time. Crucially, however, \cref{eq:logPsi} has been derived under the assumption that $w\q(\lambda,t)$ becomes stationary. To see why this assumption may be delicate, in this Appendix we consider the simpler case of interacting particles in uniform space (i.e.~without the comb constraint), such as the ones considered in e.g.~Refs.~\cite{Demery2014,venturelli2024universal} --- and corresponding to the model described in \cref{sec:model}, but with $\hat \mu(\vb x)=\hat \mu_0(\vb x)=\mu \mathbb{1}_d$, and $U_0(\vb x) = U(\vb x)$.
First, using Stratonovich calculus one can derive the exact relation~\cite{venturelli2024universal}
\begin{align}
    \partial_t  \Psi(\bm \lambda,t) = \lambda^2\mu T  -h \mu  \bm \lambda \cdot \int \dslash{q} \mathrm i \vb q \, u\q  w\q(\bm \lambda,t),
    \label{appeq:dt_psi}
\end{align}
which is analogous to \cref{eq:dt_varphi} but for $\Psi(\bm \lambda,t) = \ln\expval{\mathrm e^{\bm \lambda \cdot \vb r_0(t)}}$, and with the generating function
\begin{equation}
    w(\vb x,\bm \lambda,t) \equiv \frac{\expval{  \phi(\vb x +\vb r_0(t),t)\, \mathrm e^{ \bm \lambda \cdot \vb r_0(t)}}}{ \expval{\mathrm e^{ \bm \lambda \cdot \vb r_0(t)}} } .
\end{equation}
Following analogous steps to the ones we presented in \ref{app:wq}, one finds in the stationary state
\begin{equation}
    w_{\vb q}(\bm \lambda) = \frac{-h u\q }{T+u\q} \left[1+\frac{Ti\vb q\cdot \bm \lambda}{q^2 (2T+u\q)-2Ti\vb q\cdot \bm \lambda}\right]+\order{h^2}, \label{eq:wq_sol}
\end{equation}
where $q=\abs{\vb q}$.
From \cref{appeq:dt_psi}, this
would imply
\begin{equation}
    \Psi(\bm \lambda,t) =\mu t \left[ \lambda^2 T  -h \bm \lambda \cdot \int \dslash{q} \mathrm i \vb q\, u\q w\q(\bm \lambda)\right].
    \label{eq:CGF_continuum}
\end{equation}
By inserting the prediction for $w\q(\bm \lambda)$ found in \cref{eq:wq_sol}, we can obtain estimates of the cumulants of $\vb r_0(t)$ up to and including $\order{h^2}$, upon taking derivatives with respect to $\bm \lambda$. For instance, the fourth cumulant would read
\begin{equation}
    \frac{\expval{r_0^4(t)}_c}{t}
    =96\, h^2 \mu \,T^3  \int \dslash{q} \frac{ u\q^2 q_1^4}{(T+u\q)(2T+u\q)^3q^6} +\order{h^4},
    \label{eq:X4_continuum}
\end{equation}
where we denoted $r_0=\vu{e}_1\cdot \vb r_0 $, while $\expval{\bullet}_c$ is the connected correlation function.
However, we note that in $d\leq 2$ this integral requires regularization via the introduction of a lower (infrared) cutoff $q\geq 2\pi/L$, implying that the fourth
cumulant depends
explicitly on the size $L $ of the system --- in particular, the integral in \cref{eq:X4_continuum} becomes divergent for $L \to \infty$, signaling the breakdown of our perturbative estimate.

In fact, our prediction in \cref{eq:CGF_continuum} has to be compared with Eq.~(33) in Ref.~\cite{Demery2023}, where 
a cumulant generating function analogous to $\Psi(\bm \lambda,t)$ has been computed perturbatively for an analogous system, and for a \textit{generic} time $t$ (i.e.~not necessarily in the stationary state). In particular, the fourth cumulant reported in Eq.~(52) of Ref.~\cite{Demery2023} coincides with the one in \cref{eq:X4_continuum} only for $d>2$, whereas for $d\leq 2$ the scaling of $\expval{r_0^4(t)}_c$ with $t$ turns out to be faster than linear.

We conclude that, in general, higher-order cumulants of $r_0(t)$ only scale as $\propto t$ provided that the integral that defines the prefactor is convergent~\cite{demerypath}, which must (somewhat inconveniently) be checked a posteriori. By contrast, the integral defining the diffusion coefficient turns out to be always finite, no matter the spatial dimension $d$, both in uniform space~\cite{Demery2014} and for the case of the comb analyzed here (see \cref{sec:MGF}).

\section{Brownian dynamics simulations}
\label{app:simulation}

In this Appendix we describe the Brownian dynamics simulation of the system of interacting particles introduced in \cref{sec:model}. To this end, we focus on dimension $d=2$, and call $\vb r_i(t) = (X_i(t),Y_i(t))$.
When trying to simulate the system on a delta-like comb such as the one introduced in \cref{sec:model}, the first obvious problem is that one never has $Y_i(t)=0$ exactly, hence the particles would never actually diffuse 
horizontally.
However, there are well-established strategies in the literature to circumvent this issue: for instance, in Ref.~\cite{Ribeiro2014} the backbone is represented by a strip of spatial width $\varepsilon_c$, while in Refs.~\cite{Trajanovski_2023,Domazetoski_2020,Sandev2021} the $\delta(y)$ function that appears in the mobility matrix~\eqref{eq:mobilities2d} is smoothed out and replaced by a Gaussian. The latter strategy is the one we chose to adopt here: in practice, we simulate
\begin{align}
    \dot{X}_i(t)&=- \mu\, \cor A \left(Y_i(t)\right) \sum_{j\neq i} \pdv{X_i} U\left(\vb r_i(t)-\vb r_j(t)\right)+\sqrt{2\mu T \cor A(Y_i(t))}\eta_i^x(t),
    \label{appeq:Xi}\\
    \dot{Y}_i(t)&=- \mu\,  \sum_{j\neq i} \pdv{Y_i} U\left(\vb r_i(t)-\vb r_j(t)\right)+\sqrt{2\mu T }\eta_i^y(t),
    \label{appeq:Yi}
\end{align}
where the $\eta_i$ are uncorrelated random variables with
\begin{equation}
    \left\langle \eta_i^x(t)\eta_j^x(t')\right\rangle=\delta_{i j} \delta\left(t-t^{\prime}\right) ,
\end{equation}
and similarly for $\eta_i^y(t)$. The function $\cor A (y)$ can in principle be any smooth representation of a nascent delta function, such as
\begin{equation}
    \cor A(y) = \frac{1}{\sqrt{2\pi}\varepsilon_c}\exp(-\frac{y^2}{2\varepsilon_c^2}).
\end{equation}
Choosing an integration time step $\Delta t$ such that $\sqrt{2\mu T \Delta t} \lesssim 2\varepsilon_c$ is in general sufficient to make the measurements independent of $\varepsilon_c$ \cite{Trajanovski_2023}.
(We verified this by running several simulations with decreasing values of $\varepsilon_c$, until we found that the $\varepsilon_c$-dependence had actually been lost.)
Note that \cref{appeq:Xi,appeq:Yi} are multivariate stochastic differential equations with multiplicative noise, that we chose to integrate using a simple Euler-Maruyama algorithm (indeed, higher-order methods would not significantly speed up the numerical integration in such cases~\cite{Greiner1988,klauder1985numerical}).

In particular, choosing Gaussian interparticle potentials $U(r)= \epsilon \, \mathrm e^{-\frac{r^2}{2R^2}}$, the force terms in \cref{appeq:Xi,appeq:Yi}
reduce to
\begin{align}
    &\pdv{U(\norm{\vb r_i-\vb r_j})}{X_i}=  -\frac{X_i-X_j}{R^2}U(\norm{\vb r_i-\vb r_j}),\\ &\pdv{U(\norm{\vb r_i-\vb r_j})}{Y_i}=  -\frac{Y_i-Y_j}{R^2}U(\norm{\vb r_i-\vb r_j}) .
\end{align}
We start the simulation by preparing a system with 200 particles at density $\rho_0=1$, and fix all other parameters to unity apart from the interaction strength $\epsilon$, which we vary within the interval $\epsilon \in [0,0.5]$.
After a thermalization time $T_\T{term}=10^5$ with a time step $\Delta t=10^{-2}$, we record the trajectory of the tracer $r_0(t)$ from time $t=0$ up to time $T_\T{max}=4\cdot 10^5$, and repeat the whole process $n=10$ times. This way we end up with a collection of trajectories $\{r_0\i (t)\}_{i=1}^n$. We then use each of these trajectories to compute the time-averaged mean squared displacement~\cite{Barkai_2008,muzzeddu2024selfdiffusionanomaliesoddtracer}
\begin{equation}
    \overline{\delta^2_i(\tau, T_\T{max})} = \frac{1}{T_\T{max}-\tau}  \int_{0}^{T_\T{max} -\tau}
    \mathrm{d}t\ 
    |r_0^{(i)}(t + \tau) - r_0^{(i)}(t)|^2.
\end{equation}
Since our system is ergodic, we can 
then ensemble-average over the $n$ independent trajectories 
to obtain the estimate for the MSD 
\begin{align}
    \left\langle |r_0(t)-r_0(0)|^2\right\rangle=  \lim_{T_\T{max}\to\infty} \frac{1}{n} \sum_{i=1}^{n} 
    \overline{\delta^2_i(\tau=t, T_\T{max})},
\end{align}
whose linear fit finally leads to the estimates of the diffusion coefficient $\left\langle |r_0(t)-r_0(0)|^2\right\rangle \simeq 2 D_\text{eff} t$ reported in \cref{fig:tracer}(b).

\section*{References}

\bibliographystyle{iopart-num}
\bibliography{references} 

\end{document}